\documentclass[floatfix, preprint, onecolumn, showpacs, showkeys, preprintnumbers, nofootinbib, superscriptaddress]{revtex4-1}
\usepackage{times}
\usepackage{amssymb,amsbsy,amsmath,amsfonts}
\usepackage{graphicx}
\usepackage{float}
\usepackage{color}
\usepackage{morefloats}
\usepackage{rotating}
\usepackage{srcltx}
\usepackage{slashed}
\usepackage{multirow}
\usepackage{verbatim}
\usepackage{hyperref}
 \usepackage{ulem}

\usepackage{array}
\newcommand{\PreserveBackslash}[1]{\let\temp=\\#1\let\\=\temp}
\newcolumntype{C}[1]{>{\PreserveBackslash\centering}p{#1}}
\newcolumntype{R}[1]{>{\PreserveBackslash\raggedleft}p{#1}}
\newcolumntype{L}[1]{>{\PreserveBackslash\raggedright}p{#1}}

\begin{document}

\title{Strangeness $S=-2$ baryon-baryon interactions in relativistic chiral effective field theory}

\author{Kai-Wen Li}
\affiliation{School of Physics and Nuclear Energy Engineering and International Research Center for Nuclei and Particles
in the Cosmos, Beihang University, Beijing 100191, China}
\affiliation{Yukawa Institute for Theoretical Physics, Kyoto University, Kyoto 606-8502, Japan}

\author{Tetsuo Hyodo}
\affiliation{Yukawa Institute for Theoretical Physics, Kyoto University, Kyoto 606-8502, Japan}

\author{Li-Sheng Geng}
\email[E-mail me at: ]{lisheng.geng@buaa.edu.cn}
\affiliation{School of Physics and Nuclear Energy Engineering and International Research Center for Nuclei and Particles
in the Cosmos, Beihang University, Beijing 100191, China}
\affiliation{Beijing Key Laboratory of Advanced Nuclear Materials and Physics, Beihang University, Beijing 100191, China}

\begin{abstract}

We study the strangeness $S=-2$ baryon-baryon interactions  in relativistic chiral effective field theory at leading order. Among
the 15 relevant low energy constants,  eight of them
are determined by fitting to the state of the art  lattice QCD data of the HAL QCD Collaboration (with $m_\pi=146$ MeV), and
the rest are either taken from the  study of the $S=-1$ hyperon-nucleon systems, assuming strict SU(3) flavor symmetry, or temporarily set equal to zero.
Using the so-obtained low energy constants, we extrapolate the results  to the physical point, and show that they are consistent with
the available experimental scattering data.
 Furthermore, we demonstrate that the $\Lambda\Lambda$ and $\Xi N$ phase shifts near the $\Xi N$ threshold are very sensitive to
the lattice QCD data fitted, to the pion mass, and to isospin symmetry breaking effects. As a result, any conclusion drawn from lattice QCD data at unphysical pion masses (even close to the physical point)
should be taken with caution. Our results at the physical point,  similar to  the lattice QCD data, show that a resonance/quasi-bound state may appear in the $I=0$ $\Lambda\Lambda$/$\Xi N$ channel.
\end{abstract}


\date{\today}

\maketitle
\section{Introduction}

The strangeness $S=-2$ hyperon-nucleon ($YN$) and hyperon-hyperon ($YY$)  interactions  play a key role in many studies of great interest in hypernuclear physics and nuclear astrophysics, e.g.,  the existence of the H-dibaryon and
$\Xi$ hypernuclei, and the hyperon puzzle. Despite of the large amount of  experimental and theoretical efforts, the existence of the H-dibaryon remains inconclusive, see e.g. Refs.~\cite{Ahn:1998fj,Sasaki:2018mzh}. The H-dibaryon was first predicted to exist by Jaffe using the MIT bag model~\cite{Jaffe:1976yi}  as a deeply bound six-quark state with strangeness $S=-2$, isospin $I=0$ and spin-parity $J^P=0^+$, appearing  in the $^1S_0$ partial wave of the $\Lambda\Lambda-\Xi N-\Sigma\Sigma$ coupled channels. Recent lattice QCD simulations performed at $m_\pi\gtrsim 389$ MeV showed some  evidence
for the existence of a bound H-dibaryon below the $\Lambda\Lambda$ threshold~\cite{Beane:2010hg,Inoue:2010es,Beane:2011iw,Francis:2018qch}. However, subsequent studies showed that when those results are extrapolated to the physical region  the H-dibaryon  becomes either weakly bound or unbound~\cite{Shanahan:2011su,Haidenbauer:2011ah,Inoue:2011ai,Yamaguchi:2016kxa}. Lately the HAL QCD Collaboration performed simulations very close to the physical region~\cite{Sasaki:2018mzh}, namely, $m_\pi=146$ MeV. Using the so-called HAL QCD method and assuming SU(3) flavor symmetry,
they obtained an effective $\Lambda\Lambda-\Xi N$ coupled-channel potential. The calculations
using such a potential yielded a resonant state in the $\Lambda\Lambda$ channel (a quasi-bound state in the $\Xi N$ channel), 
which, however, show sizable systematic uncertainties, depending on the evolution time $t$ in their simulation. Furthermore, it was shown that the coupled-channel effects between $\Lambda\Lambda$ and $\Xi N$ are weak. 

Regarding the existence of $\Xi$ hypernuclei~\cite{Yamaguchi:2001ip,Friedman:2007zza,Hiyama:2010zz}, a moderately attractive interaction is inferred from the $^{12}C(K^-,K^+)^{12}_\Xi$Be reaction~\cite{Khaustov:1999bz}. However, subsequent analyses showed that the $\Xi$ potential could be either attractive~\cite{Khaustov:1999bz}, almost vanishing~\cite{Kohno:2009ny} or weakly repulsive~\cite{Krishichayan:2010zza}. In 2015, the ``KISO" event claimed a deeply bound $\Xi^--^{14}$N hypernucleus~\cite{Nakazawa:2015joa}, indicating at least an attractive $\Xi N$ interaction. On the other hand, based on the few-body calculations of the $\Xi NN$ hypernucleus~\cite{Garcilazo:2015noa}, a $\Xi NN$ bound state might appear, indicating that  the $\Xi N$ interaction might be strongly attractive. 

$YN$ and $YY$ interactions are important inputs to astrophysical studies  as well, since hyperons might appear in the interior region of neutron stars. The inclusion of $YN$ interactions results in a softening of the equation-of-state (EoS) of nuclear matter, which is inconsistent with the observations of  two-solar-mass neutron stars~\cite{Demorest:2010bx,Antoniadis:2013pzd}, known as the ``hyperon puzzle". In this case, repulsive $YY$ interactions seem to provide one possible solution by stiffening the EoS~\cite{Lonardoni:2013gta,Maslov:2015msa}.

In this work, we study the strangeness $S=-2$ $YN$ and $YY$ interactions in relativistic chiral effective field theory (ChEFT) at leading order (LO). It is an extension of our previous studies of the nucleon-nucleon ($NN$)~\cite{Ren:2016jna,Ren:2017yvw} and strangeness $S=-1$ $YN$~\cite{Li:2016paq,Li:2016mln,Li:2017zwn,Song:2018qqm} systems. The relativistic ChEFT has been shown to be able to describe the $NN$, $\Lambda N$ and $\Sigma N$ scattering data fairly well, already at LO~\cite{Ren:2016jna,Ren:2017yvw,Li:2016paq,Li:2016mln,Li:2017zwn,Song:2018qqm,Ren:2018xxd,Ren:2017thl}. In contrast to the $S=0$ and $S=-1$ sectors, there are only a few experimental data in the $S=-2$ sector. Here we will use the latest lattice QCD data of
 the HAL QCD Collaboration~\cite{Sasaki:2018mzh} to fix eight of the 15 low energy constants (LECs) at LO. The rest are determined from the $S=-1$ sector~\cite{Li:2016mln} assuming SU(3) flavor symmetry, or temporarily set equal to zero. In addition, we will extrapolate the results to the physical region and compare them with the available $\Lambda\Lambda$ and $\Xi N$ scattering data. The consistency between the lattice QCD
 simulations, ChEFT and experimental data will be discussed.

The paper is organized as follows: In Sec.~\ref{II} we present a brief overview of the formalism of relativistic ChEFT. The fits to the lattice QCD data are discussed in Sec.~\ref{III}. Phase shifts,  cross sections and low energy parameters for the $\Lambda\Lambda$, $\Sigma\Sigma$,  and $\Xi N$ systems are shown in Sec.~\ref{IV}. We conclude with a short summary and outlook in Sec.~\ref{V}.

\section{Baryon-baryon interactions in relativistic chiral effective field theory}\label{II}

ChEFT has been successfully applied to study low-energy (octet) baryon-baryon interactions~\cite{Epelbaum:2008ga,Machleidt:2011zz,Haidenbauer:2013oca,Polinder:2007mp,Haidenbauer:2015zqb,Haidenbauer:2009qn} since the pioneering work of  Weinberg~\cite{Weinberg:1990rz,Weinberg:1991um}.  Compared to phenomenological models, ChEFT has three main advantages. First,
it has a deep connection with the underlying theory of the strong interactions,  QCD, particularly, chiral symmetry and its breaking.
Second, it employs a  power counting scheme, which enables one to improve calculations systematically. As a result,
one can estimate the uncertainty of the results. In addition, multi-baryon forces can be treated on the same footing as
the two-body interactions. Recently, we explored a relativistic ChEFT approach to study the $NN$~\cite{Ren:2016jna} and $YN$~\cite{Li:2016mln} interactions at LO, in which more relativistic effects are taken into account in the potentials and scattering equation than in the non-relativistic ChEFT.

The main feature of the relativistic formalism is that the complete  baryon spinors are retained in the calculations.
\begin{equation}\label{ub}
  u_B(\mbox{\boldmath $p$}, s)= N_p
  \left(
  \begin{array}{c}
    1 \\
    \frac{\mbox{\boldmath $\sigma$}\cdot \mbox{\boldmath $p$}}{E_p+M_B}
  \end{array}\right)
  \chi_s,
  ~~~~~~~~
  N_p=\sqrt{\frac{E_p+M_B}{2M_B}},
\end{equation}
where $E_p=\sqrt{\mbox{\boldmath $p$}^2+M_B^2}$, and $M_B$ is the averaged baryon mass. Apparently, Lorentz invariance is maintained by such a treatment. Details of the formalism can be found in Refs.~\cite{Ren:2016jna,Li:2016mln}.

For the strangeness $S=-2$ sector, the LO potentials consist of non-derivative four-baryon contact terms (CT) and one-pseudoscalar-meson exchanges (OPME). 15 independent LECs appear in the CT that have to be pinned down by fitting to either experimental or lattice QCD data. Strict SU(3) symmetry is imposed on the CT and the coefficients of OPME, which can be found in e.g. Refs.~\cite{Polinder:2007mp,Haidenbauer:2015zqb}. However, due to the mass difference of the exchanged mesons ($\pi,K,\eta$), SU(3) symmetry is broken in the OPME. We have followed the convention of Ref.~\cite{Polinder:2007mp} and our previous $S=-1$ work~\cite{Li:2016mln} to redefine the LECs such as $C^{\Lambda\Lambda}_{1S0}$, instead of using the SU(3) representation such as $C^{27}_{1S0}$. In addition to the 12 LECs already appearing in the $S=-1$ sector~\cite{Li:2016mln}, 3 more (independent) LECs, defined as
\begin{align}
	& V_{\rm{CT}}^{\Lambda\Lambda\rightarrow\Lambda\Lambda}(^1S_0)
	=
	\xi_B \left[ C_{1S0}^{4\Lambda}(1+R_p^2R_{p'}^2) + \hat C_{1S0}^{4\Lambda}(R_p^2+R_{p'}^2) \right], \\
	& V^{\Lambda\Lambda\rightarrow\Lambda\Lambda}_{\textrm{CT}}(^3P_1) 
	= 
    \xi_B \left( -\frac{4}{3}C_{3P1}^{4\Lambda}R_pR_{p'} \right),
\end{align}
appear in the $S=-2$ sector. Here $\xi_B=N_p^2N_{p'}^2$, $R_p=|\mbox{\boldmath $p$}|/(E_p+M_B)$, $R_{p'}=|\mbox{\boldmath $p$}'|/(E_{p'}+M_B)$. 
To obtain the scattering amplitude $T_{\rho\rho'}^{\nu\nu',J}$, the coupled-channel Kadyshevsky equation is solved,
\begin{align}\label{SEK}
  & T_{\rho\rho'}^{\nu\nu',J}(p',p;\sqrt{s})
  =
   V_{\rho\rho'}^{\nu\nu',J}(p',p)
   \notag\\
    &\qquad
   +
  \sum_{\rho'',\nu''}\int_0^\infty \frac{dp''p''^2}{(2\pi)^3} \frac{M_{B_{1,\nu''}}M_{B_{2,\nu''}}~ V_{\rho\rho''}^{\nu\nu'',J}(p',p'')~
   T_{\rho''\rho'}^{\nu''\nu',J}(p'',p;\sqrt{s})}{E_{1,\nu''}E_{2,\nu''}
  \left(\sqrt{s}-E_{1,\nu''}-E_{2,\nu''}+i\epsilon\right)},
\end{align}
where $V_{\rho\rho'}^{\nu\nu',J}$ is the interaction kernel which consists of CT and OPME, $\sqrt{s}$ is the total energy of the baryon-baryon system in the center-of-mass frame and $E_{n,\nu''}=\sqrt{\mbox{\boldmath $p$}''^2+M_{B_{n,\nu''}}^2}$, $(n=1,2)$. The labels $\rho,\rho',\rho''$ denote the partial waves, and $\nu,\nu',\nu''$ denote the particle channels. The Coulomb interaction is not considered in the present work due to the lack of near-threshold data and because it would require a complicated treatment. This is  consistent with the lattice QCD simulations~\cite{Sasaki:2018mzh}. In order to avoid ultraviolet divergence in solving the scattering equation, the chiral potentials are multiplied by an exponential form factor,
\begin{equation}\label{EF}
  f_{\Lambda_F}(p,p') = \exp \left[-\left(\frac{p}{\Lambda_F}\right)^{4}-\left(\frac{p'}{\Lambda_F}\right)^{4}\right] \, ,
\end{equation}
 with a cutoff value $\Lambda_F=600$ MeV.\footnote{We have chosen the value of $\Lambda_F$ that can best describe the strangeness $S=-1$ $YN$ scattering data~\cite{Li:2016mln}, though
 acceptable fits to the data can be obtained with a cutoff ranging from 550 to 800 MeV (see Ref.~\cite{Song:2018qqm} for more discussions).}

\section{A fit to the lattice QCD results}\label{III}

Recently the HAL QCD Collaboration performed simulations for the strangeness $S=-2$ baryon-baryon systems with almost physical pion masses ($m_\pi= 146$ MeV)~\cite{Sasaki:2018mzh}. The so-called  HAL QCD approach is employed to extract the potentials from the Nambu-Bethe-Salpeter wave functions on the lattice. Although the resulting potentials should in principle be independent of the measured time slice $t$, current results shows sizable dependence on the evolution time $t$, which should be regarded as the systematic uncertainty of the lattice QCD simulation~\cite{private}.
 They obtained results for the $I=2$ $\Sigma\Sigma$ $^1S_0$ phase shifts, the  $I=0$ $\Xi N$ $^3S_1$ phase shifts~\cite{private}, the  $I=0$ $\Lambda\Lambda$, and $\Xi N$ $^1S_0$ phase shifts and the inelasticity~\cite{Sasaki:2018mzh} using the effective $\Lambda\Lambda-\Xi N$ coupled channels, instead of the full $\Lambda\Lambda-\Xi N-\Sigma\Sigma$ coupled channels. 

In the present work, we fit these lattice QCD data~\cite{Sasaki:2018mzh,private} to determine the relevant eight LECs of the CT. The fits are performed in the following steps.\footnote{The relevant
masses and coupling constants are fixed at $m_\pi=146$ MeV, $m_K=525$ MeV, $m_N=958$ MeV, $m_\Lambda=1140$ MeV, $m_\Sigma=1223$ MeV, and $m_\Xi=1354$ MeV~\cite{Sasaki:2018mzh}. In addition, we have used $D+F=g_A=1.277$, $F/(F+D)=0.4$ and $f_0\simeq f_\pi=92.2$ MeV~\cite{Li:2016mln}. } 

First, we fitted to the lattice QCD  $I=2$ $\Sigma\Sigma$ $^1S_0$ phase shifts with the center-of-mass energy $E_{\rm{cm}}\leq 40$ MeV, where $E_{\rm{cm}} = \sqrt{s} - M_{B_1} - M_{B_2}$. $M_{B_1}$ and $M_{B_2}$ are the baryon masses of the channel with the lowest energy threshold. This is a single-channel scattering and the two LECs $C^{\Sigma\Sigma}_{1S0}$ and $\hat C^{\Sigma\Sigma}_{1S0}$ can be fixed. All results with $t=11-13$ were used to estimate the central value and the uncertainty of the phase shift at each energy. 

Second, the $^3S_1$ partial wave of the $I=0$ $\Xi N$ system is treated in the same way. Note that in our convention the relevant LECs are defined as,
\begin{align}
	V_{\rm{CT} ,I=0}^{\Xi N\rightarrow\Xi N}(^3S_1)
	&=
	\xi_B \left[ \frac{1}{9}\big(C_{3S1}^{\Lambda\Lambda}-C_{3S1}^{\Lambda\Sigma}\big)(9+R_p^2R_{p'}^2) + \frac{1}{3}\big(\hat C_{3S1}^{\Lambda\Lambda}-\hat C_{3S1}^{\Lambda\Sigma}\big)(R_p^2+R_{p'}^2) \right] \notag\\
	&= \xi_B \left[ \frac{1}{9}C^{8a}_{3S1}(9+R_p^2R_{p'}^2) + \frac{1}{3}\hat C^{8a}_{3S1}(R_p^2+R_{p'}^2) \right].
\end{align}
In this case only the two combinations of those four relevant LECs can be pinned down, namely $C^{8a}_{3S1}$ and $\hat C^{8a}_{3S1}$. 
For the LECs (or the combinations of LECs) that contribute to the SU(3) structure $10$ and $10^*$ in the $^3S_1$ partial waves, we have taken their values from the $S=-1$ sector via SU(3) symmetry~\cite{Li:2016mln}.

Six LECs appear in the spin-singlet $\Lambda\Lambda-\Xi N-\Sigma\Sigma$ coupled channels, $C^{\Sigma\Sigma}_{1S0}$, $\hat C^{\Sigma\Sigma}_{1S0}$, $C^{\Lambda\Lambda}_{1S0}$, $\hat C^{\Lambda\Lambda}_{1S0}$, $C^{4\Lambda}_{1S0}$ and $\hat C^{4\Lambda}_{1S0}$, but two of them $C^{\Sigma\Sigma}_{1S0},\hat C^{\Sigma\Sigma}_{1S0}$ have been fixed from the $I=2$ $\Sigma\Sigma$ $^1S_0$ phase shifts as described above. Unlike the $I=2$ $\Sigma\Sigma$ $^1 S_0$ and $I=0$ $\Xi N$ $^3 S_1$ cases, the lattice QCD data on the $I=0$ $\Lambda\Lambda$ and $\Xi N$ phase shifts obtained at various time $t$ look rather different. A resonant $\Lambda\Lambda$ state (a quasi-bound $\Xi N$ state below the threshold) is found with the $t=9,10,11$ lattice QCD data, but not with the $t=12$ data. Therefore, we have performed separate fits to the lattice QCD data obtained at different $t$ ranging form $9$ to $12$. The low energy $\Lambda\Lambda$ $^1S_0$ phase shifts with $E_{\rm{cm}}\leq 20$ MeV, the $\Lambda\Lambda$, $\Xi N$ phase shifts, and the inelasticity with $32 \leq E_{\rm{cm}}\leq 32.8$ MeV are taken into account. Because the $\Xi N$ quasi-bound state appears very close to the $\Xi N$ threshold at $E_{\rm{cm}}=32$ MeV with $m_{\pi}=146$ MeV, the near-threshold data are included. 

We summarize the details of the lattice QCD data used and the corresponding LECs in Table~\ref{lqcddata}.  The values of the $S$-wave LECs are listed in Table~\ref{LECs}. The LEC $C_{3P1}^{4\Lambda}$ in the $^3P_1$ partial wave of the $\Lambda\Lambda\rightarrow\Lambda\Lambda$ reaction is not determined by this analysis, but it contributes to the $\Lambda\Lambda$ and $\Xi^- p$ induced cross sections. We temporarily set $C_{3P1}^{4\Lambda}=0$ for the calculation of the cross section, assuming that the low-energy cross section is dominated by the $S$-wave contribution.

\begin{table}[h]
\footnotesize
\centering
 \caption{Lattice QCD data  used in the fits and the corresponding independent LECs of the relativistic ChEFT approach.}\label{lqcddata}
 \begin{tabular}{llccl}
  \hline
  \hline
    Reaction~~~~ & $~~~I~$ & ~~~Partial wave ~~~& Phase shifts 
    & Corresponding LECs  \\
  \hline
  $\Sigma\Sigma\rightarrow\Sigma\Sigma$
   & $~~~2$ & $^1S_0$
   & $E_{\rm{cm}}\leq 40$ MeV~\cite{private} 
   & $C^{\Sigma\Sigma}_{1S0}$, $\hat C^{\Sigma\Sigma}_{1S0}$ \\
   \hline
  $\Xi N\rightarrow\Xi N$
   & $~~~0$& $^3S_1$
   & $E_{\rm{cm}}\leq 40$ MeV~\cite{private} 
   & $C^{8a}_{3S1}$, $\hat C^{8a}_{3S1}$ 
   \\
   \hline
  $\Lambda\Lambda\rightarrow\Lambda\Lambda$
   & $~~~0$& $^1S_0$
   & $E_{\rm{cm}}\leq 20$ MeV, $32$ MeV $\leq E_{\rm{cm}}\leq 32.8$ MeV~\cite{Sasaki:2018mzh} 
   & 
     \\
   $\Xi N\rightarrow\Xi N$
   & $~~~0$& $^1S_0$
   & $32 \leq E_{\rm{cm}}\leq 32.8$ MeV~\cite{Sasaki:2018mzh} 
   &  $C^{\Lambda\Lambda}_{1S0}$, $\hat C^{\Lambda\Lambda}_{1S0}$,
      $C^{4\Lambda}_{1S0}$, $\hat C^{4\Lambda}_{1S0}$ 
     \\
   Inelasticity
   & $~~~0$& $^1S_0$
   & $32 \leq E_{\rm{cm}}\leq 32.8$ MeV~\cite{Sasaki:2018mzh} 
   & 
     \\
  \hline
  \hline
\end{tabular}
 \end{table}

\begin{table}[h]
\centering
 \caption{LECs for the $S$-wave contact terms (in unit of $10^4$ GeV$^{-2}$). The $^{3}S_{1}$ LECs are decomposed with the help of the $S=-1$ scattering data~\cite{Li:2016mln}, assuming SU(3) symmetry.}\label{LECs}
 \begin{tabular}{ccccccc}
  \hline
  \hline
   ~~~~~~\phantom{$C^{\Sigma\Sigma}_{1S0}$}~~~~~~
   &~~~~~~ $C^{\Sigma\Sigma}_{1S0}$ ~~~~~~&~~~~~~ $\hat C^{\Sigma\Sigma}_{1S0}$ ~~~~~
   &~~~~~~ $C^{\Lambda\Lambda}_{1S0}$ ~~~~~~&~~~~~~ $\hat C^{\Lambda\Lambda}_{1S0}$ ~~~~~~
   &~~~~~~ $C^{4\Lambda}_{1S0}$ ~~~~~~&~~~~~~ $\hat C^{4\Lambda}_{1S0}$~~~~~~ \\
   \hline
           & $-0.0418$ & ~~~$0.1726$ & & & & \\\hline
  
   $t=9$~~ &  &  & $-0.0154$   & ~~~$0.0041$ & $-0.0088$   & ~~~$0.3570$ \\
   $t=10$  &  &  & $-0.0183$   & ~~~$0.0977$ & $-0.0134$   & ~~~$0.6544$ \\
   $t=11$  &  &  & $-0.0202$   & $-0.0482$   & $-0.0038$   & ~~~$0.8982$ \\
   $t=12$  &  &  & ~~~$0.0157$ & ~~~$0.6119$ & ~~~$0.1709$ & $-0.1982$ \\
  \hline
  \hline
   & $C^{\Lambda\Lambda}_{3S1}$ & $\hat C^{\Lambda\Lambda}_{3S1}$
   & $C^{\Sigma\Sigma}_{3S1}$ & $\hat C^{\Sigma\Sigma}_{3S1}$
   & $C^{\Lambda\Sigma}_{3S1}$ & $\hat C^{\Lambda\Sigma}_{3S1}$ \\
  \hline
   & ~~~$0.0137$ & ~~~$0.9261$ & ~~~$0.0872$ & $-0.4132$ & ~~~$0.0230$ & ~~~$0.2880$  \\
  \hline
  \hline
\end{tabular}
 \end{table}

\section{Results and discussion}\label{IV}

\subsection{The $I=2$ $\Sigma\Sigma$ $^1S_0$ phase shifts}\label{IVA}

\begin{figure}[h]
 \centering
	\includegraphics[width=0.8\textwidth,bb=0 0 289 215]{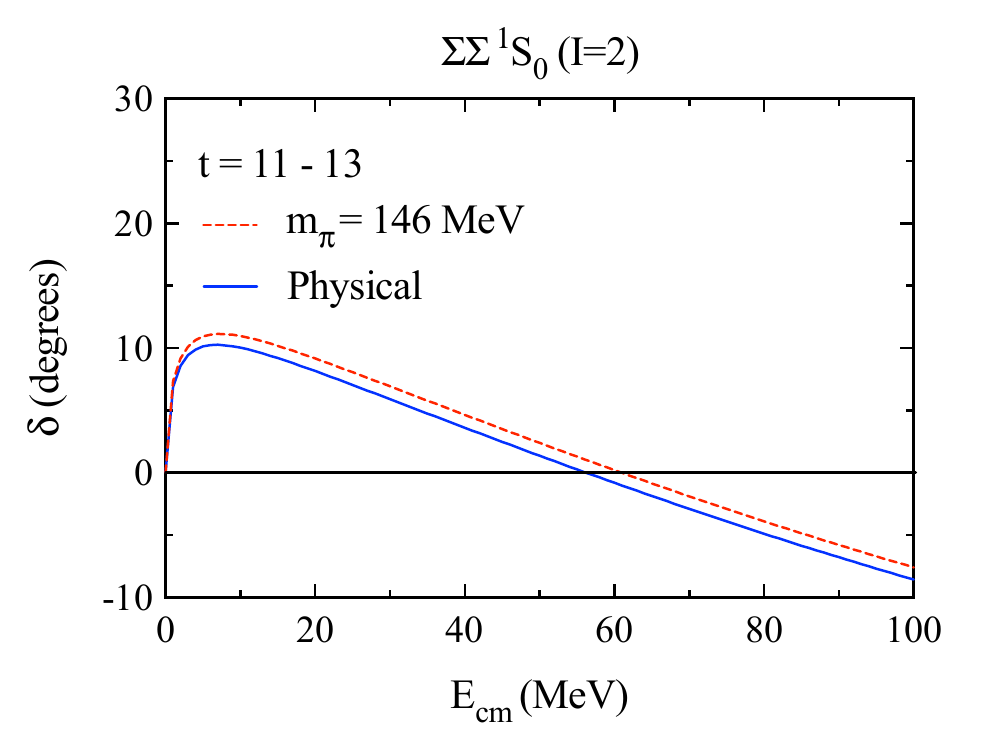}
	\caption{Phase shifts of  the $I=2$ $\Sigma\Sigma$ $^1S_0$ partial wave. The dashed 
	line denotes
	the 
	result  with $m_\pi=146$ MeV and the solid 
	line denotes
	the 
	result with
	the physical pion mass.}\label{PSSiSi}
\end{figure}

In Fig.~\ref{PSSiSi}, we show the $I=2$ $\Sigma\Sigma$ $^1S_0$ phase shifts. The dashed lines are the fitted results with $m_\pi=146$ MeV.
We  obtained a $\chi^2/d.o.f.=0.08$ after the fits, which indicates a good description of the lattice QCD data. The solid lines are the extrapolations to the physical pion mass, with the isospin symmetry being assumed for the hadron masses. The extrapolations were done by only changing the hadron masses to their physical values, but keeping the coupling constants $F$, $D$ and $f_0$ and the other LECs fixed. 

For the $\Sigma\Sigma$ $^1 S_0$ channel, the phase shifts at the low-energy region  are positive, indicating that the attractions are weak, but at the high energy region the interactions become repulsive.
In the SU(3) basis, the $^1S_0$ partial wave of $\Sigma\Sigma$ ($I=2$), $\Sigma N$ ($I=3/2$) and $NN$ ($I=1$) all belong to the same representation of 27. However, the maximum value of the phase shifts are about 10 degrees for 
the $\Sigma\Sigma$ system, 40 degrees for the $\Sigma N$ system~\cite{Li:2016mln}, and 60 degrees for the $NN$ system~\cite{Ren:2016jna}. 
This clearly tells that the 27 SU(3) representation is becoming less attractive with the increase of the strangeness. On the other hand, we checked that 
a simultaneous fit of the $\Sigma^+ p$ cross sections and the lattice $\Sigma\Sigma$ $^1S_0$ phase shifts failed, similar to the attempt at  a combined fit of $NN$ and strangeness $S=-1$ $YN$ data~\cite{Li:2016mln}. As a result, we conclude that  SU(3) breaking effects should be included if one wishes to simultaneously describe the systems with different strangeness, as also discussed in Ref.~\cite{Haidenbauer:2015zqb}. 
We note that the extrapolation to the physical point only causes minor change of the phase shift.

\subsection{The $I=0$ $\Xi N$ $^3S_1$ phase shifts}

\begin{figure}[h]
 \centering
	\includegraphics[width=0.8\textwidth,bb=0 0 289 215]{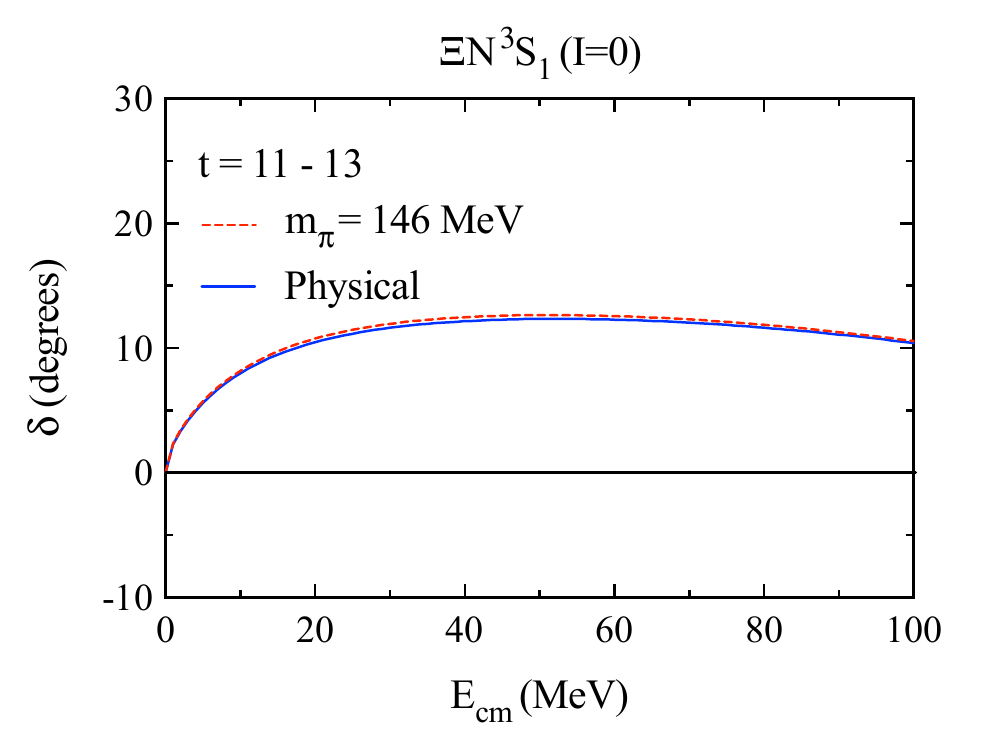}
	\caption{Phase shifts of $I=0$ $\Xi N$ $^3S_1$ partial waves. The dashed 
	line denotes the 
	result with $m_\pi=146$ MeV and the solid 
	line denotes the 
	result with
	the physical pion mass.}\label{PSXiN}
\end{figure}

The $\Xi N$ $^3S_1$ phase shifts are shown in Fig.~\ref{PSXiN}, with a fitted $\chi^2/d.o.f. = 2.68$. The relativistic ChEFT can describe the low
energy  lattice data well, but not those of high energies~\cite{private}.  Namely, lattice data show that the phase shift turns into negative at high energies, which is not reproduced in the present study.  It seems that  higher order chiral potentials are needed in this channel in order to provide enough repulsion at high energies.
In this channel, the phase shift remains almost the same after the chiral extrapolation to the physical point as well.

\subsection{The $\Lambda\Lambda-\Xi N-\Sigma\Sigma$ $^1S_0$ phase shifts}

\begin{figure}[h]
 \centering
	\includegraphics[width=0.33\textwidth,bb=0 0 291 216]{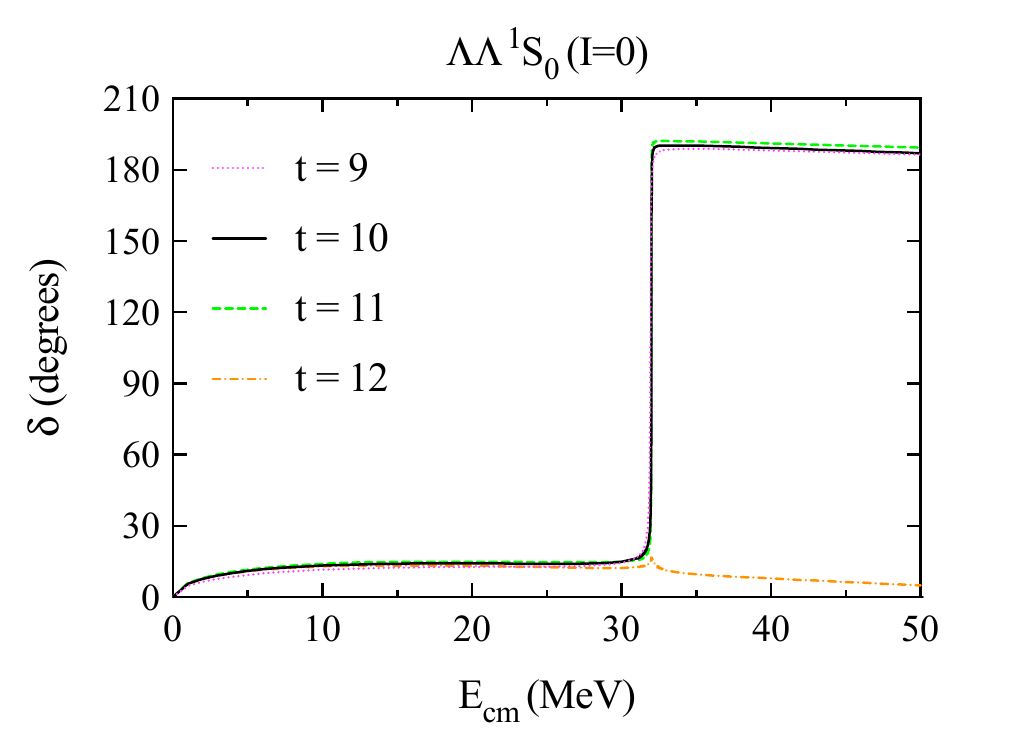}~
	\includegraphics[width=0.33\textwidth,bb=0 0 291 216]{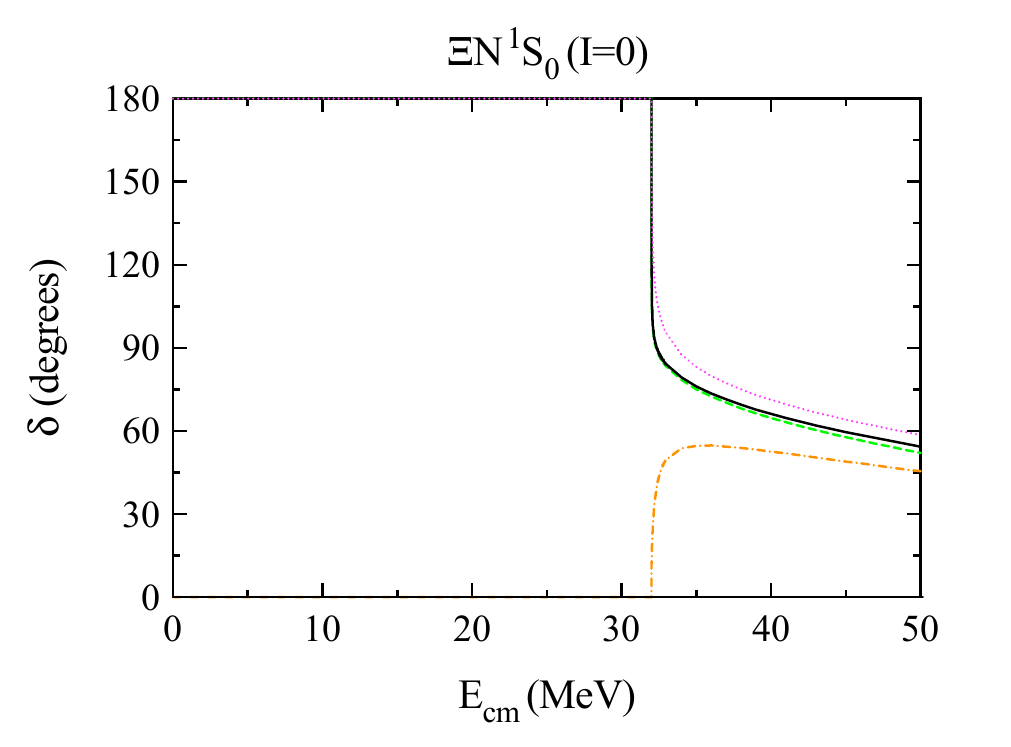}~
	\includegraphics[width=0.33\textwidth,bb=0 0 288 213]{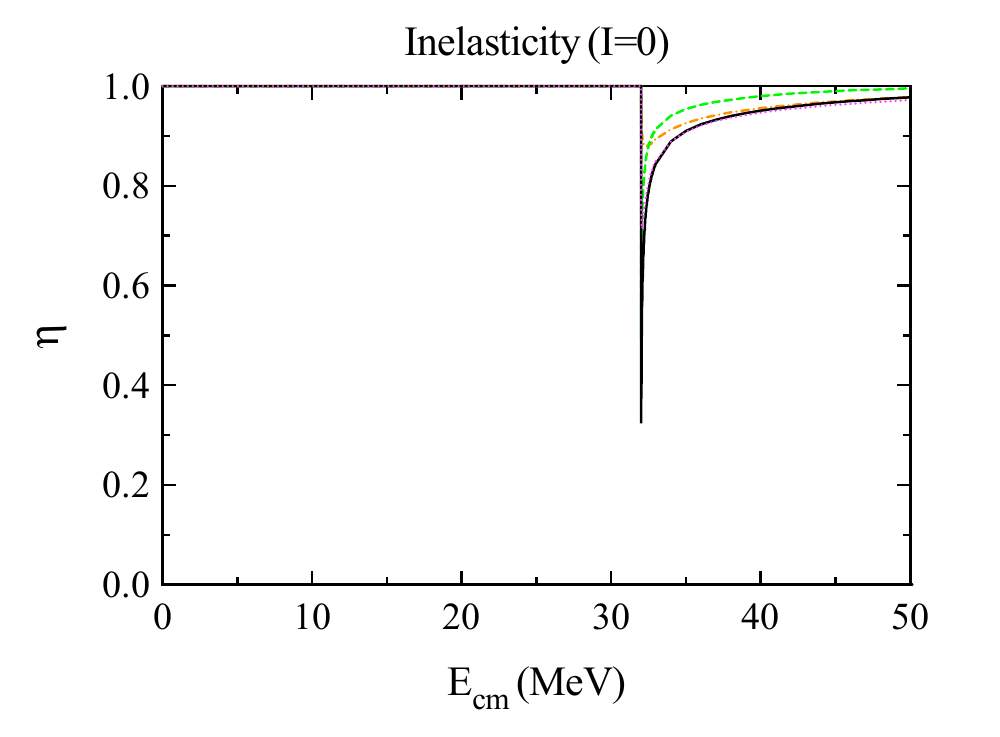}
	\caption{$I=0$ $\Lambda\Lambda$, $\Xi N$ $^1S_0$ phase shifts and the inelasticity with $m_\pi=146$ MeV and $t=9-12$. The inelasticity $\eta$ is defined as 
	$S_{ii}=\eta e^{2i\delta_{i}}$.}\label{PSLL3}
\end{figure}

\begin{figure}[h]
 \centering
	\includegraphics[width=0.33\textwidth,bb=0 0 291 218]{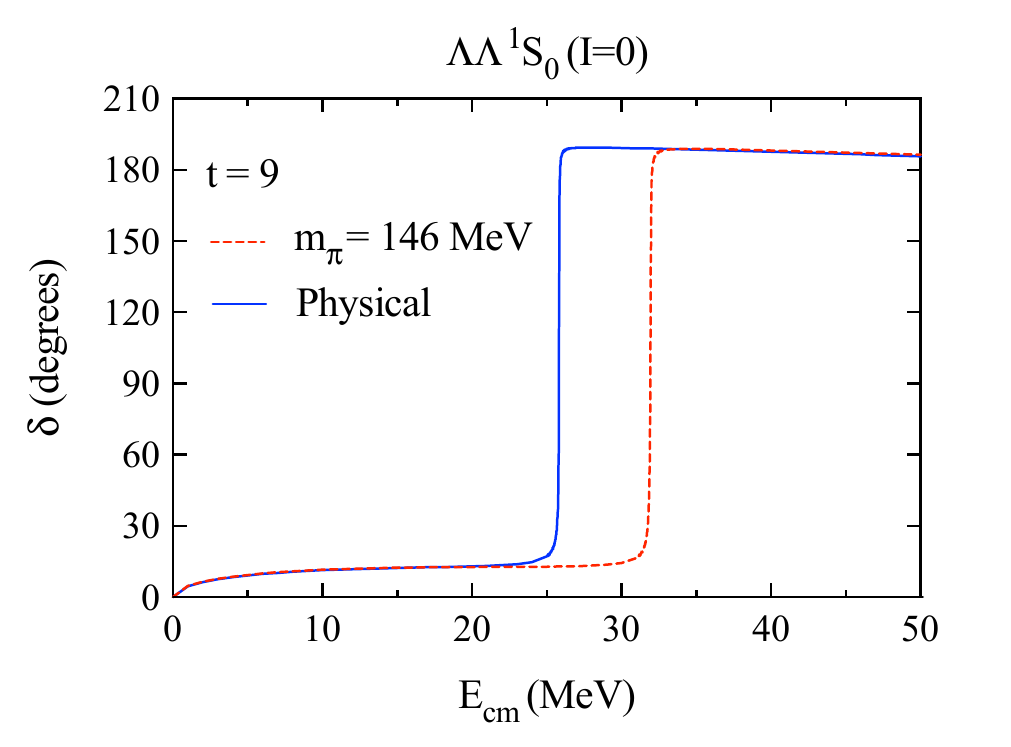}~
	\includegraphics[width=0.33\textwidth,bb=0 0 291 216]{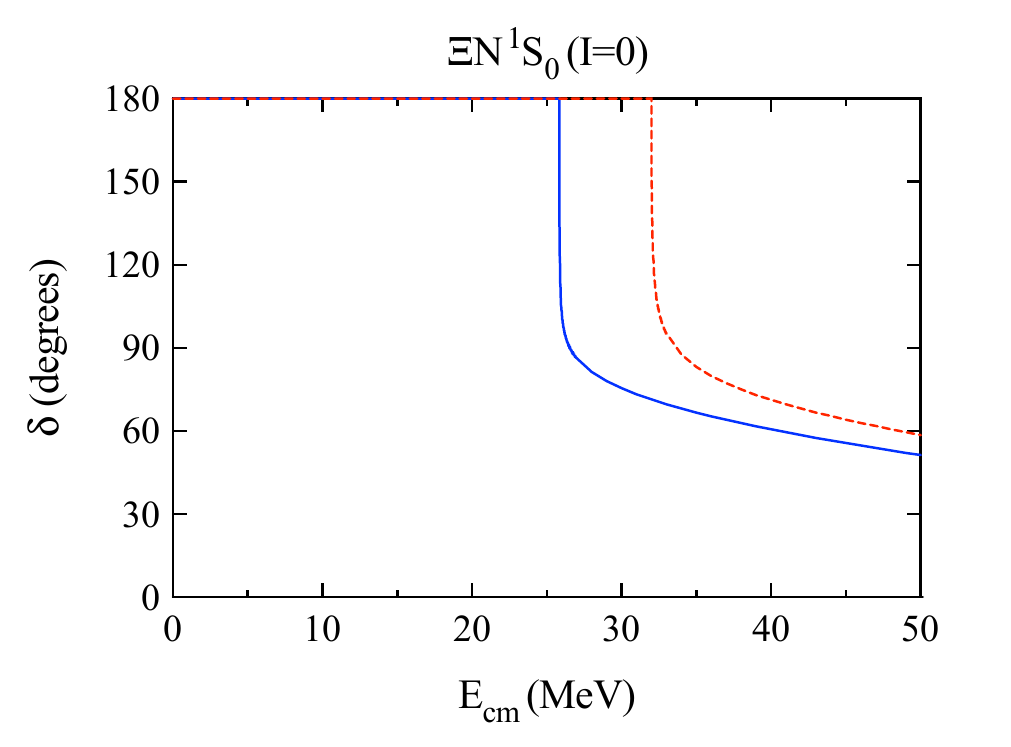}~
	\includegraphics[width=0.33\textwidth,bb=0 0 288 213]{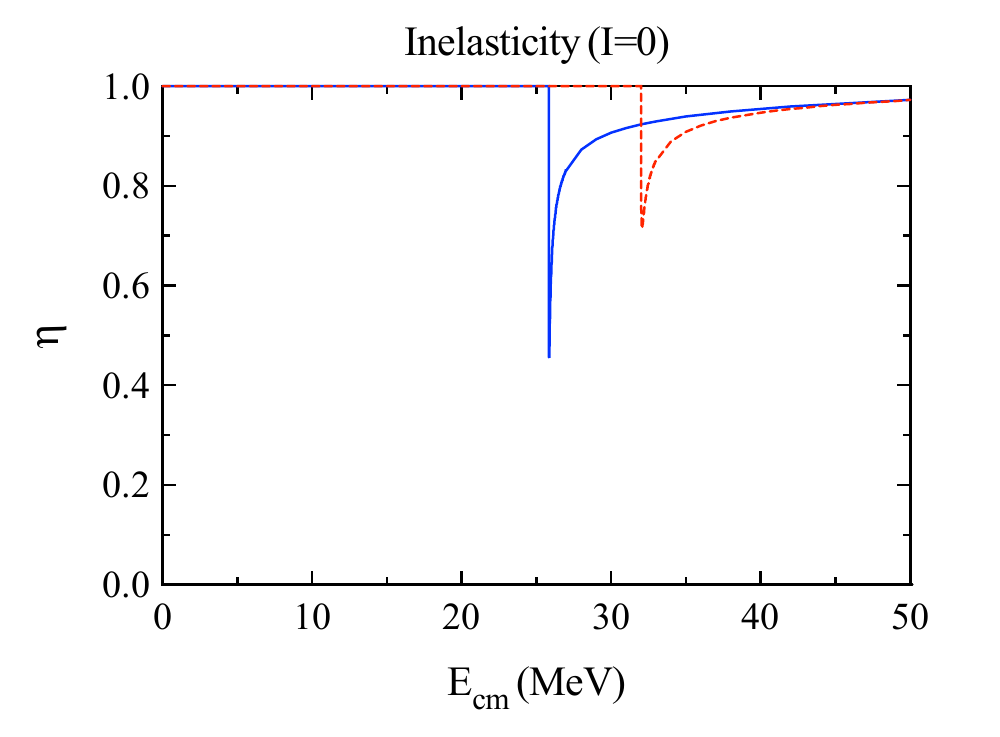} \\
	\includegraphics[width=0.33\textwidth,bb=0 0 291 218]{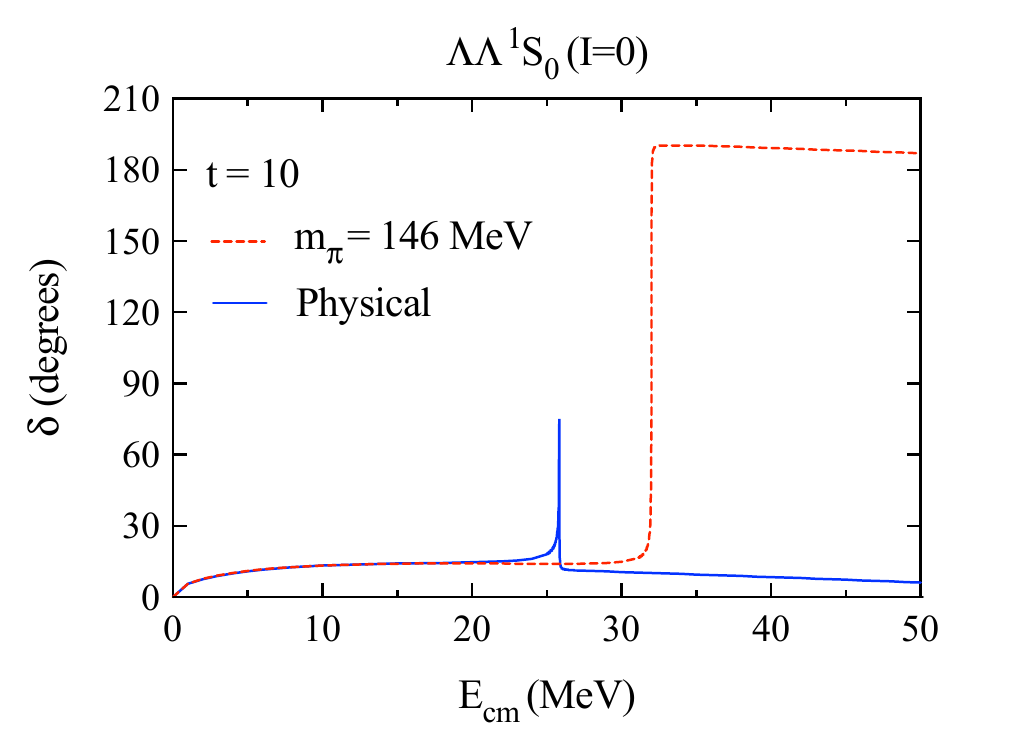}~
	\includegraphics[width=0.33\textwidth,bb=0 0 291 218]{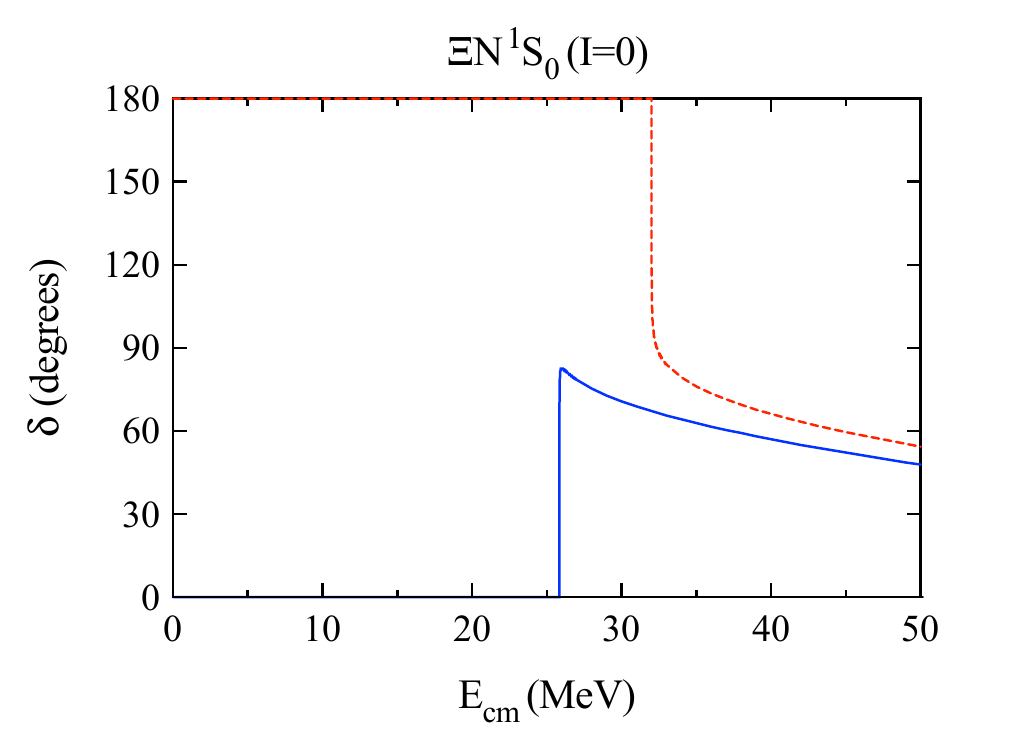}~
	\includegraphics[width=0.33\textwidth,bb=0 0 288 213]{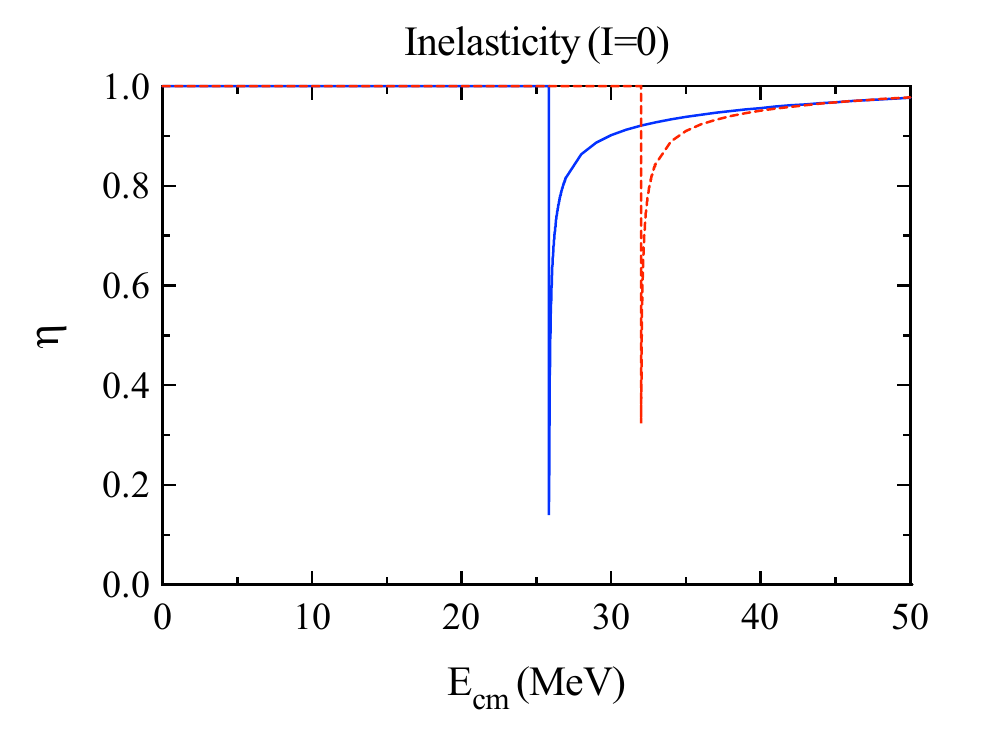} \\
	\includegraphics[width=0.33\textwidth,bb=0 0 291 218]{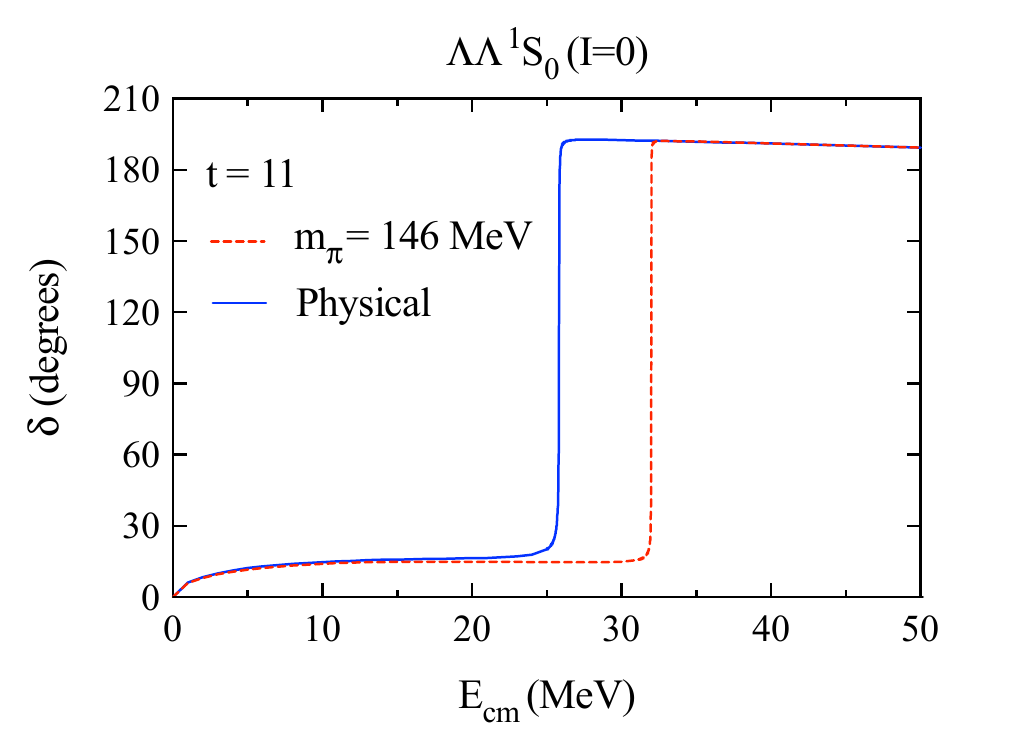}~
	\includegraphics[width=0.33\textwidth,bb=0 0 291 218]{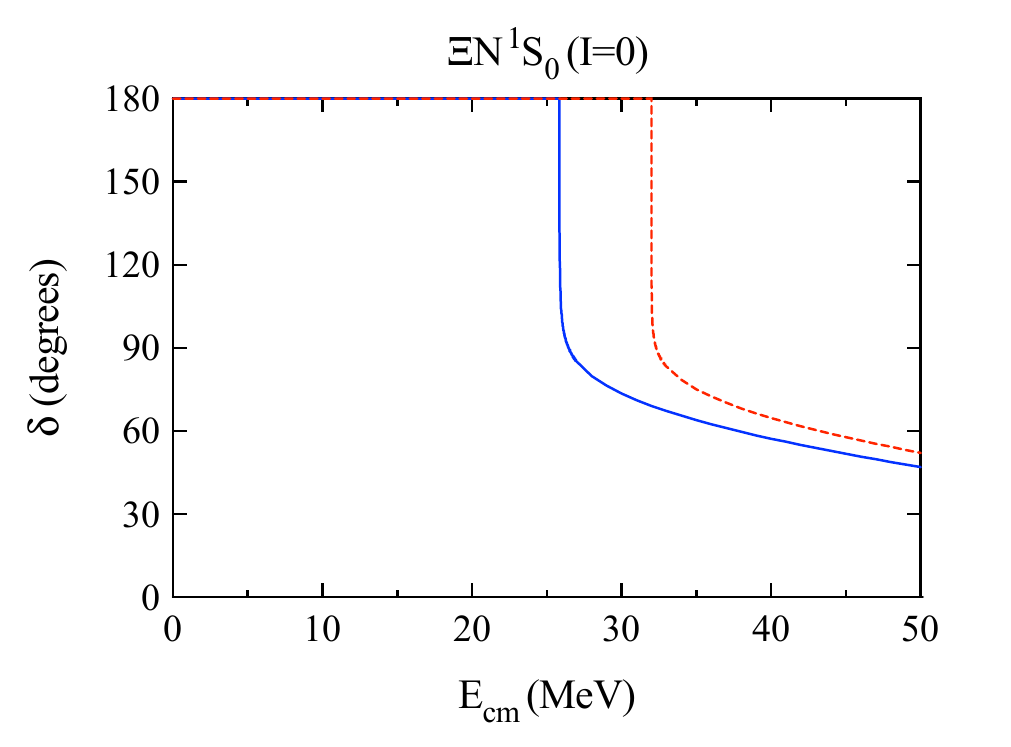}~
	\includegraphics[width=0.33\textwidth,bb=0 0 288 213]{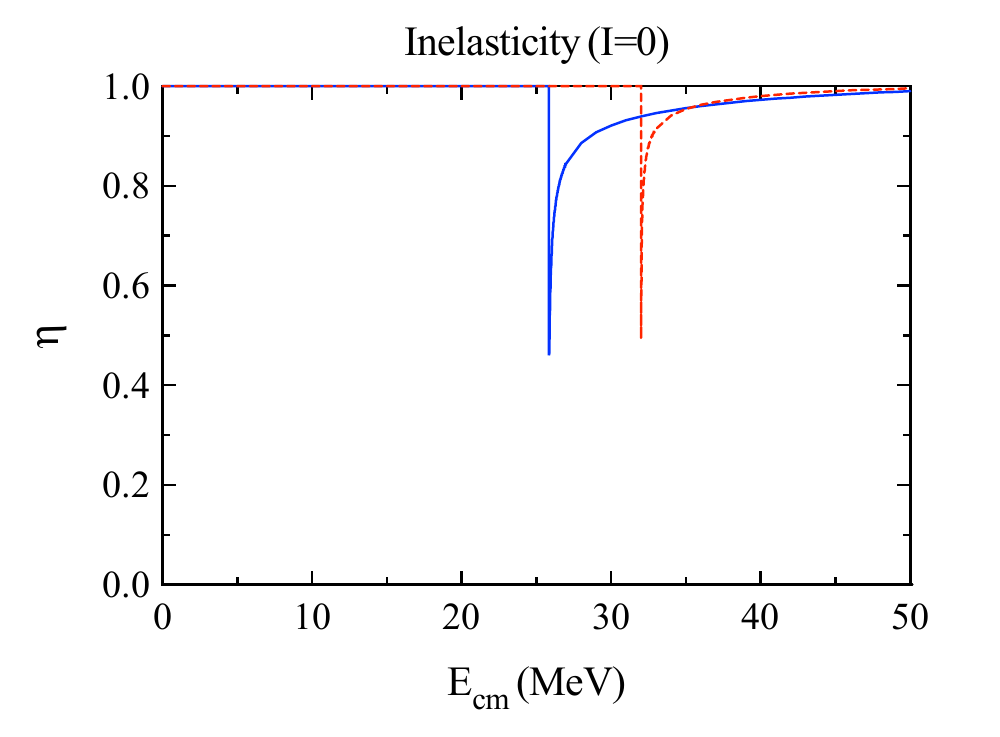} \\
	\includegraphics[width=0.33\textwidth,bb=0 0 291 218]{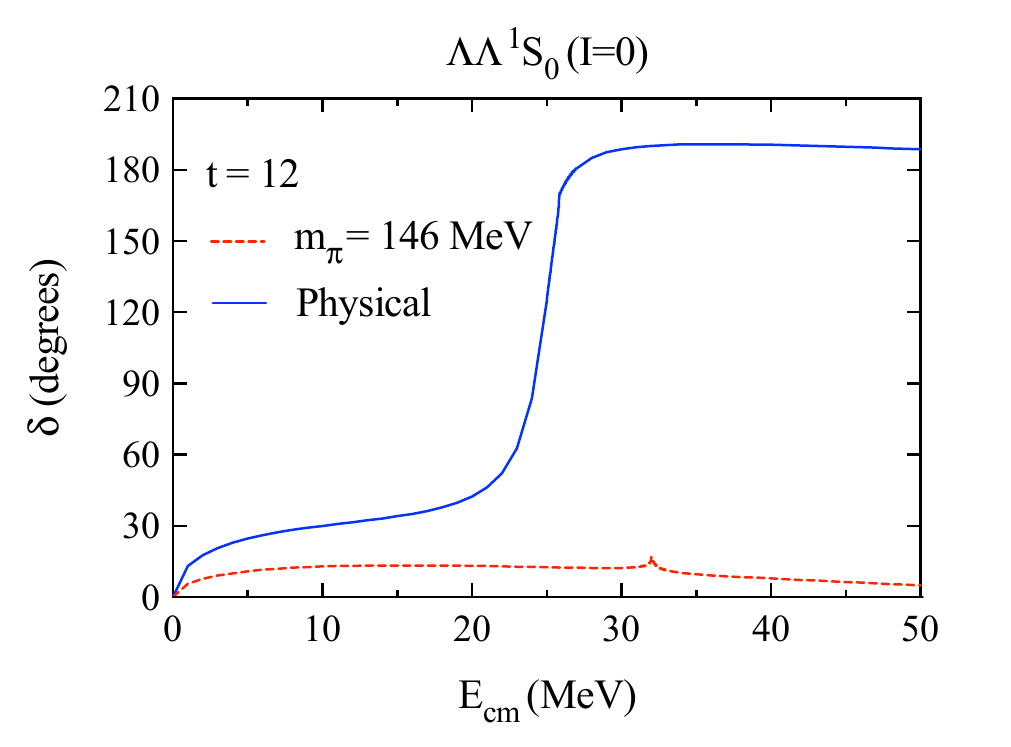}~
	\includegraphics[width=0.33\textwidth,bb=0 0 291 218]{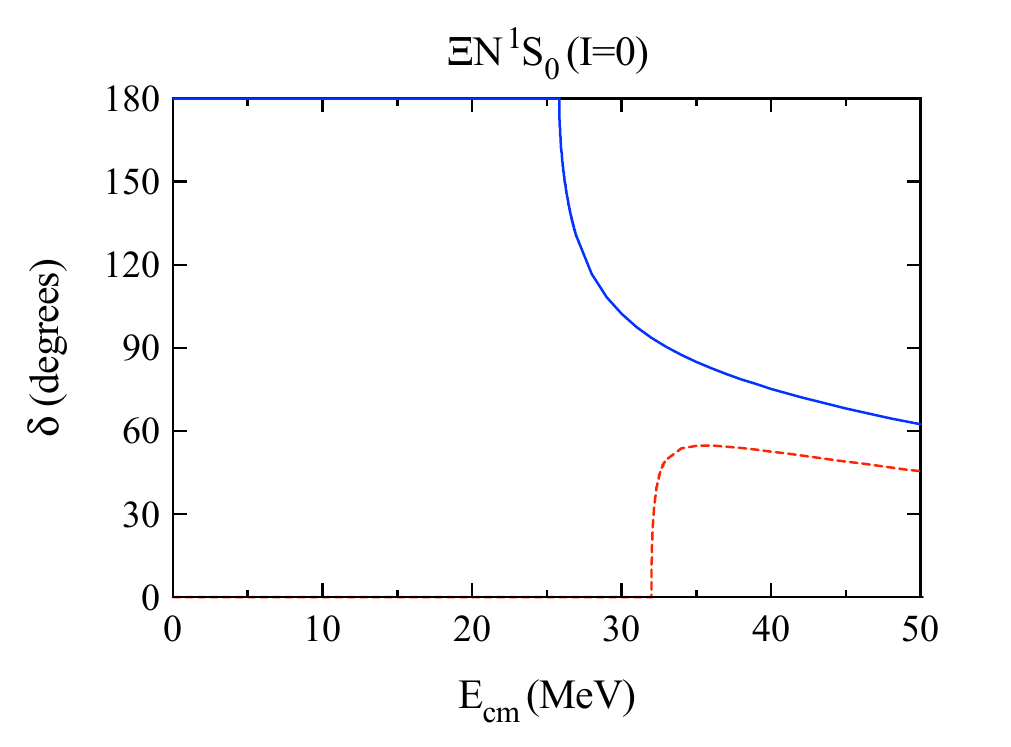}~
	\includegraphics[width=0.33\textwidth,bb=0 0 288 213]{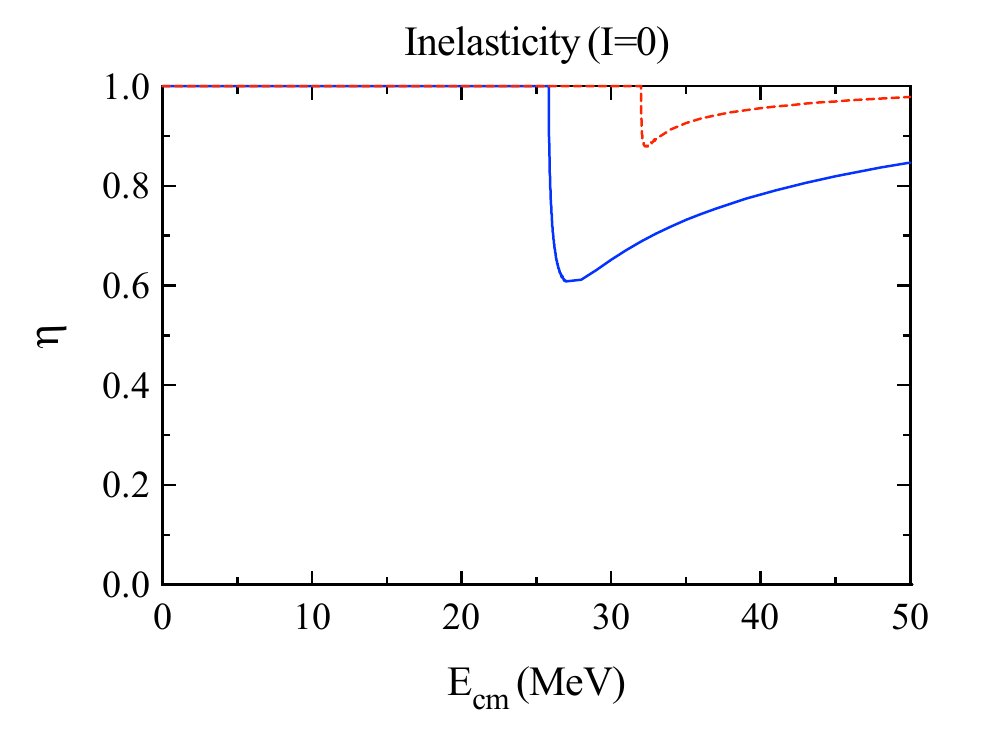}
	\caption{$I=0$ $\Lambda\Lambda$, $\Xi N$ $^1S_0$ phase shifts and the inelasticity with $m_\pi=146$ MeV (dashed lines) and at the physical 
	point (solid lines) with $t=9-12$.}\label{PSt9-12}
\end{figure}

As for the $\Lambda\Lambda-\Xi N-\Sigma\Sigma$ coupled-channel, which is important for the study of  the H-dibaryon, we can obtain a good description of the lattice QCD data on the $\Lambda\Lambda$, $\Xi N$ phase shifts and the inelasticity for each $t=9,10,11,12$, with the corresponding $\chi^2/d.o.f.=0.42,0.11,0.30,0.01$, respectively. The results are shown in Fig.~\ref{PSLL3}.  The sharp resonant state of $\Lambda\Lambda$ (the quasi-bound state of $\Xi N$) is well reproduced for $t=9-11$. However, the extrapolations to the physical pion mass
look quite different for $t=9-12$, as shown in Fig.~\ref{PSt9-12}. The sharp resonance remains with $t=9$ and $t=11$, but it disappears with $t=10$. For the case of $t=12$, a quasi-bound state appears in the $\Xi N$ system  after the extrapolation, while the quasi-bound state is absent at $m_{\pi}=146$ MeV. Note that the $\Xi N$ threshold has changed after the extrapolation, because the baryon masses changed as well.
The origin of this difference of the extrapolation will be discussed in Sec.~\ref{sec:XiNquasibound}.
We have also calculated the $\Lambda\Lambda$ scattering lengths in the physical region with $t=9-12$, and found that they are  consistent with the analyses from the hypernuclear experiments and the analysis of the two-particle correlations in the heavy ion collisions~\cite{Rijken:2006ep,Fujiwara:2006yh,Filikhin:2002wm,Afnan:2003ty,Filikhin:2004sn,Yamada:2004ks,Vidana:2003ic,Usmani:2004vs,Nemura:2004xb,Gasparyan:2011kg,Morita:2014kza,Adamczyk:2014vca}, as shown in Table~\ref{LaLaSL}.

\begin{table}[h]
\centering
 \caption{ Physical $\Lambda\Lambda$ $^{1}S_0$ scattering length with $t=9-12$  (in units of fm).}\label{LaLaSL}
 \begin{tabular}{lccccc}
  \hline
  \hline
     & ~~~~$t=9$~~~~ & ~~~~$t=10$~~~~ & ~~~~$t=11$~~~~ & ~~~~$t=12$~~~~ & Exp. analyses~\cite{Rijken:2006ep,Fujiwara:2006yh,Filikhin:2002wm,Afnan:2003ty,Filikhin:2004sn,Yamada:2004ks,Vidana:2003ic,Usmani:2004vs,Nemura:2004xb,Gasparyan:2011kg,Morita:2014kza,Adamczyk:2014vca} \\
  \hline
   $a^{\Lambda\Lambda}_{1S0}$ & $-0.49$~~ & $-0.60$~~ & $-0.67~~$ & $-1.44$~~ & $-1.87 \sim -0.5$ \\
  \hline
  \hline
\end{tabular}
 \end{table}

\subsection{The $\Xi N$ quasi-bound state}\label{sec:XiNquasibound}

\begin{figure}[h]
 \centering
	\includegraphics[width=0.8\textwidth,bb=0 0 299 212]{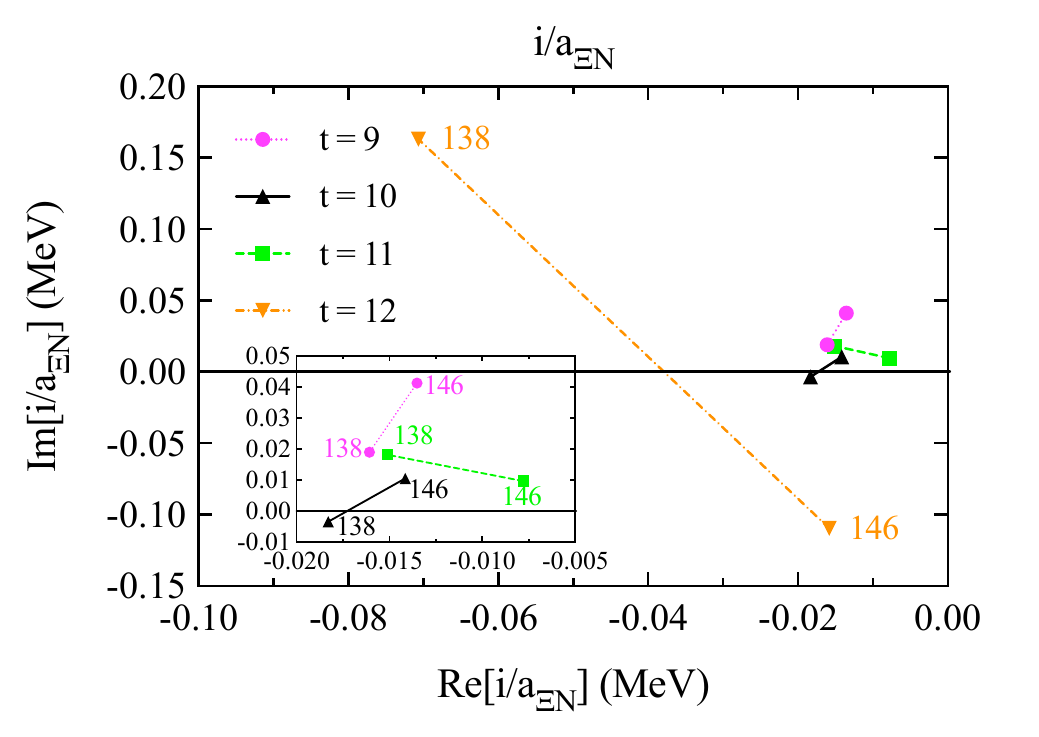}
	\caption{Inverse of the $^1S_0$ scattering length of the $I=0$ $\Xi N$ channel as a function of the pion mass. }\label{PPXiN}
\end{figure}

Our above study showed that the existence of the $\Xi N$ quasi bound sate (the H-dibaryon) is a quite delicate issue.
In order to understand the different behavior of the extrapolation, we show the inverse of the $^1S_0$ scattering length of the $\Xi N$ channel multiplied with $i$, i.e., $i/a_{\Xi N}$, in Fig.~\ref{PPXiN}. 
Because $\Xi N$ is not the coupled channel with the lowest threshold, the scattering length $a_{\Xi N}$ is in general complex due to the decay to the $\Lambda\Lambda$ channel. When $|a_{\Xi N}|$ is much larger than the typical length scale of the strong interaction $\sim$ 1 fm, $i/a_{\Xi N}$ represents approximately the pole position of the $\Xi N$ scattering amplitude in the complex momentum plane. If Im $(i/a_{\Xi N})>0$, then the pole is in the first Riemann sheet of the complex energy plane, indicating that the $\Xi N$ system has a quasi-bound state. 

One can see that for $t=9$ and $10$, the evolution from $m_\pi = 146$ MeV to the physical pion mass is similar. The value of the imaginary part decreases  and  finally becomes negative for $t=10$ when extrapolated to the physical region, which corresponds to the disappearance of the quasi-bound state in the $\Xi N$ system. Im $(i/a_{\Xi N})<0$ indicates that the pole is in the second Riemann sheet of the $\Xi N$ channel, which is not the most adjacent sheet to the physical scattering axis and hence the structure is not directly visible in  observables. While for $t=11$ and $12$, the trend is opposite. In both cases, the imaginary part of $i/a_{\Xi N}$ increases, and a quasi-bound state appears in the physical region. Especially for $t=12$, the scale of the movement is relatively larger compared with the other three cases. Such a behavior originates from the values of the LECs with $t=12$, e.g., the magnitude of $\hat C^{\Lambda\Lambda}_{1S0}$ and $C^{4\Lambda}_{1S0}$ are larger than those with $t=9-11$, as shown in Table~\ref{LECs}.
In this way, the fate of the quasi bound state in the extrapolation procedure is very sensitive. Even small change of the inverse scattering length at $m_{\pi}=146$ MeV can result in completely different behavior at the physical point.

\begin{figure}[h]
 \centering
	\includegraphics[width=0.5\textwidth,bb=0 0 291 216]{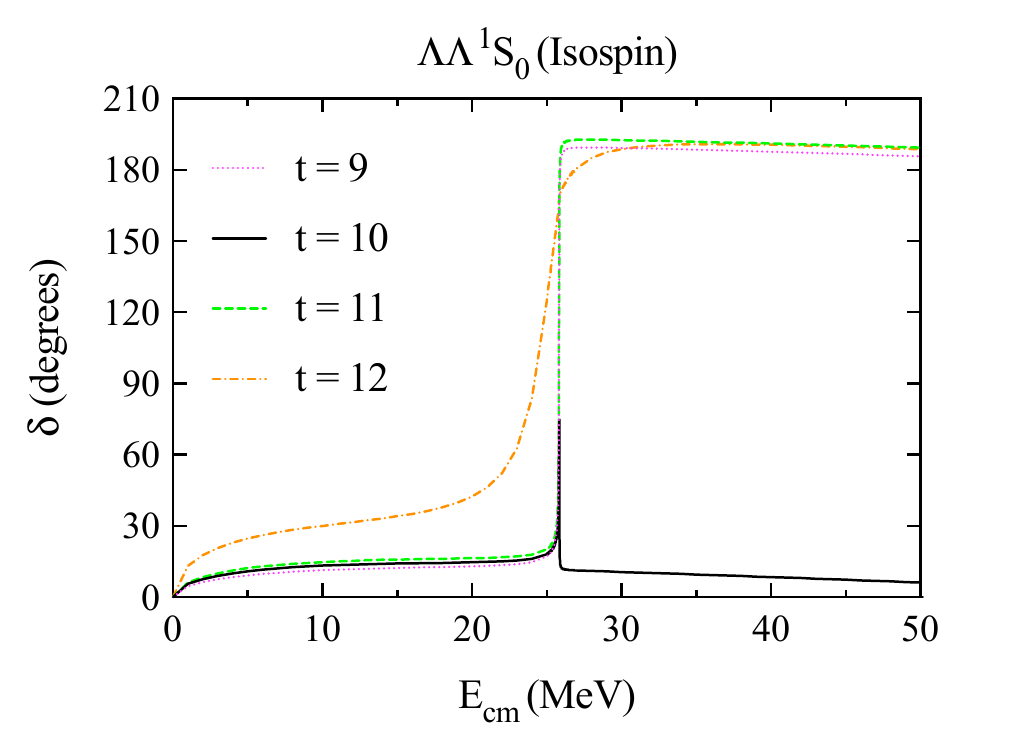}~
	\includegraphics[width=0.5\textwidth,bb=0 0 291 216]{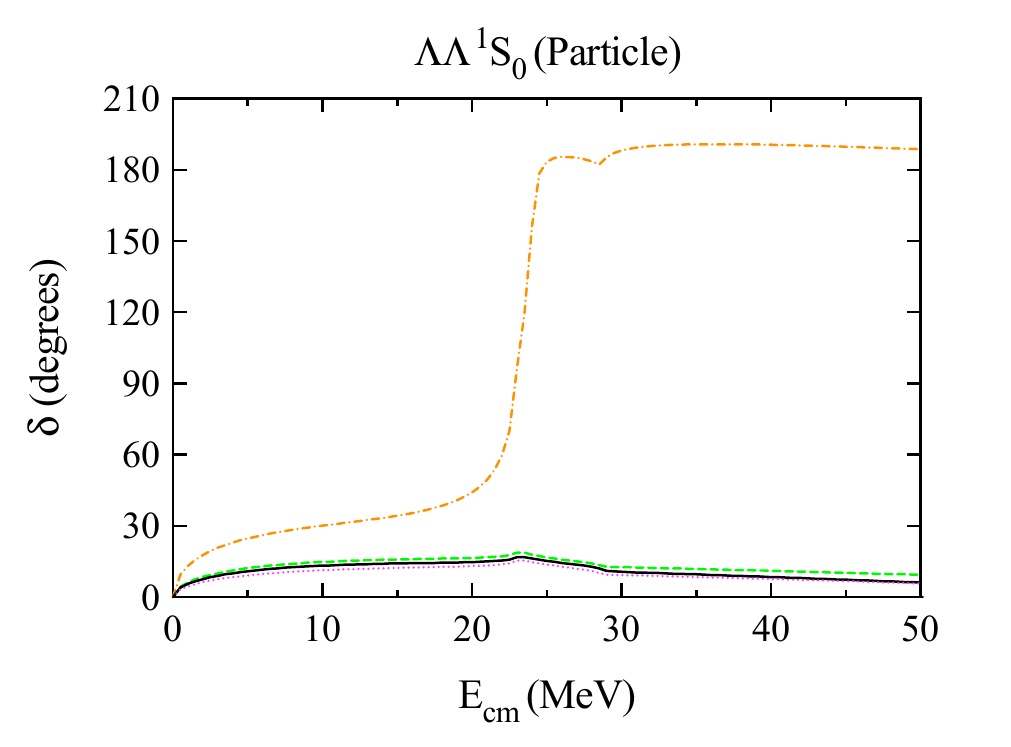}
	\caption{$\Lambda\Lambda$ $^1S_0$ phase shifts with isospin 
	averaged baryon masses (left) and with physical baryon masses (right).}\label{LaLa138}
\end{figure}

The above calculations are performed in the isospin basis, where it is a $\Lambda\Lambda-\Xi N-\Sigma\Sigma$ coupled-channel with a common baryon mass being used for each isospin multiplet. If we consider isospin breaking effects in the baryon masses, we should calculate in the $\Lambda\Lambda-\Xi^0 n-\Xi^- p-\Sigma^0\Lambda-\Sigma^0\Sigma^0-\Sigma^-\Sigma^+$ coupled channels. In Fig.~\ref{LaLa138} we compare the $\Lambda\Lambda$ $^1S_0$ phase shifts obtained with or without isospin symmetry for the baryon masses. Note that with the physical baryon masses, the threshold energy of $\Xi^0 n$ is different from that of $\Xi^- p$, and there appear two threshold cusps around $E_{\rm cm}\sim 25$ MeV. It can be seen that those sharp resonant states have disappeared if the isospin breaking effects are included. Only for the $t=12$ case  the resonant state appears at the $\Xi^0 n$ threshold, which corresponds to a quasi-bound state of the $\Xi^0 n$ system. 

We summarize the different scenarios for the existence of a $\Xi N$ bound state in Table~\ref{XiNbound}. The results are based on the fits to the central values of the lattice QCD data.  It can be seen that the quasi-bound state in the $\Xi N$ system is extremely sensitive to the lattice QCD data fitted, to the pion mass, and to the isospin breaking effects.
We note that the strong sensitivity of the behavior of the phase shift around the $\Xi N$ threshold with respect to the isospin symmetry breaking effect was also discussed in Ref.~\cite{Haidenbauer:2015zqb}. However, the $\Lambda\Lambda$ scattering length remains almost the same with or without isospin breaking effects taken into account. 

\begin{table}[h]
\footnotesize
\centering
 \caption{Summary of the $\Xi N$ quasi-bound state in different scenarios. The quasi-bound state exists for $\bigcirc$ cases. }\label{XiNbound}
 \begin{tabular}{lccccl}
  \hline
  \hline
    Lattice data & ~~~~$t=9$~~~~ & ~~~~$t=10$~~~~ & ~~~~$t=11$~~~~ & ~~~~$t=12$~~~~  \\
  \hline
  $m_{\pi}=146$ MeV
   & $\bigcirc$ & $\bigcirc$ & $\bigcirc$ & 
    \\
  $m_{\pi}=138$ MeV with isospin average baryon masses
   & $\bigcirc$ & 
   & $\bigcirc$ & $\bigcirc$ \\
  Physical hadron masses
   & 
   & 
   & 
   & $\bigcirc$ \\
  \hline
  \hline
\end{tabular}
 \end{table}

\begin{figure}[h]
 \centering
	\includegraphics[width=0.33\textwidth,bb=0 0 291 218]{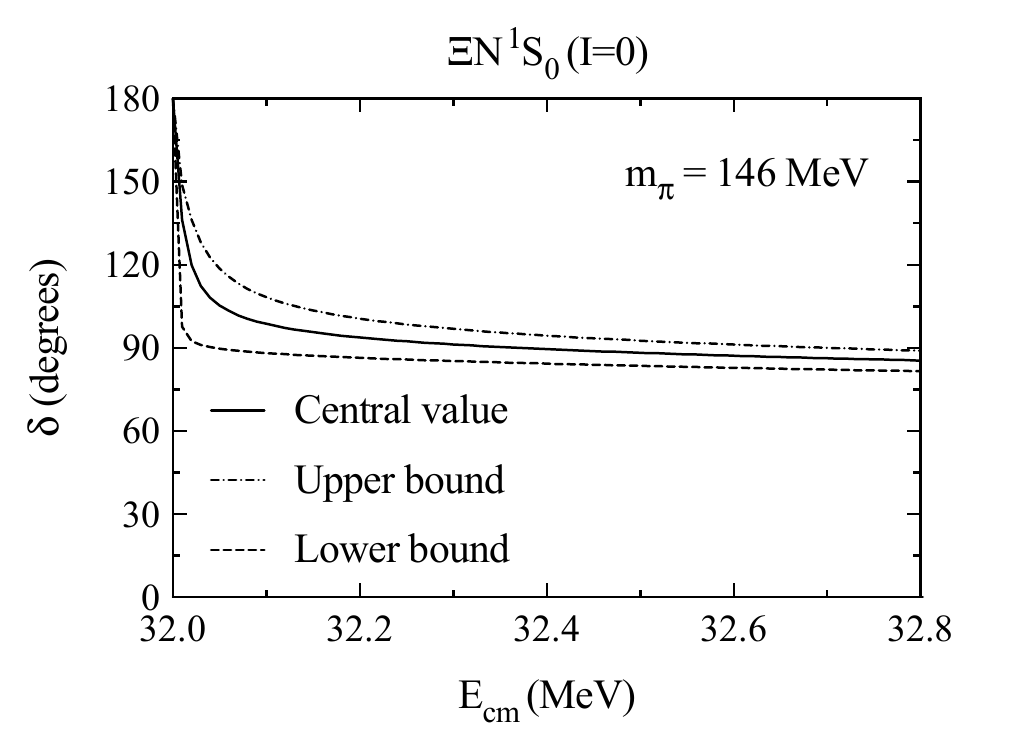}~
	\includegraphics[width=0.33\textwidth,bb=0 0 291 218]{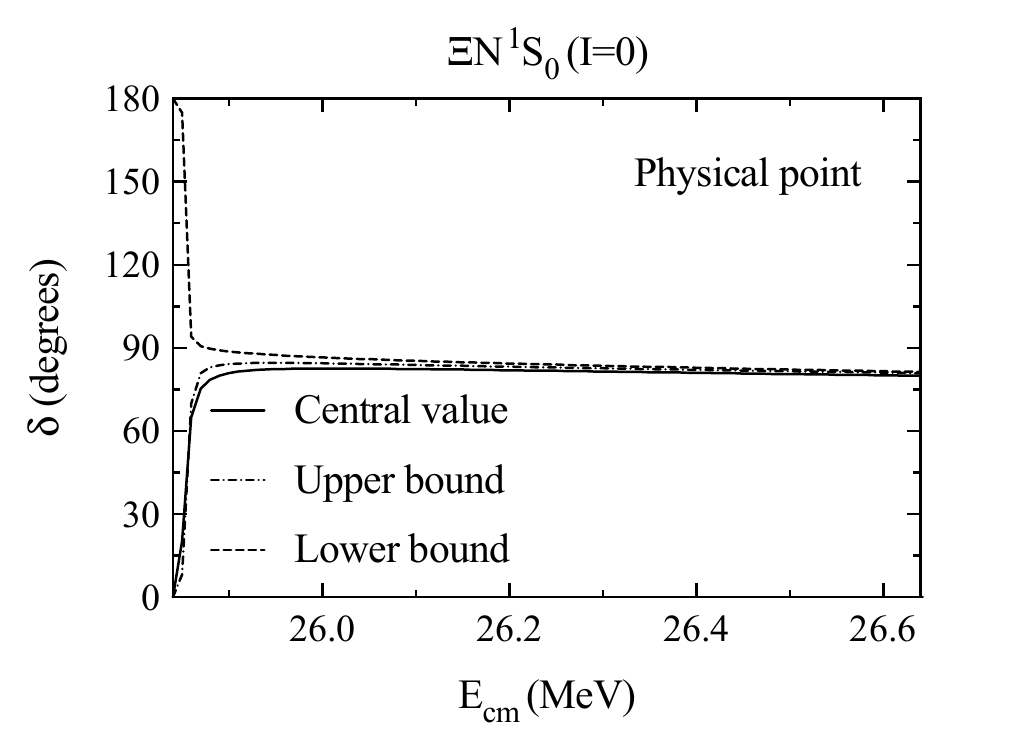}~
	\includegraphics[width=0.33\textwidth,bb=0 0 288 213]{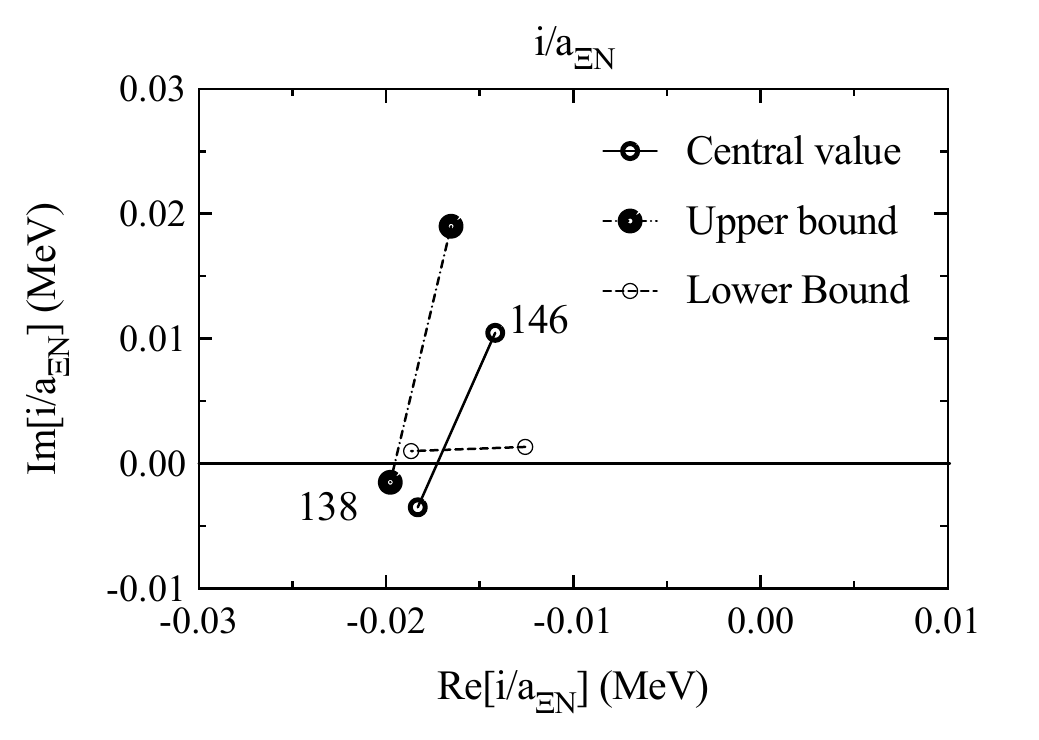}
	\caption{Near-threshold $\Xi N$ $^1S_0$ phase shifts at $m_\pi =146$ MeV (left), physical point (middle) and the extrapolations of $i/a_{\Xi N}$  (right) with $I=0$  at $t=10$ within the lattice QCD error bands.}\label{PP10Er}
\end{figure}

In our study, we have also taken into account the statistical errors of the lattice QCD results. 
In principle, the lattice QCD simulations are more reliable as the time $t$ increases, but the uncertainties increase as well. To  balance reliability and accuracy, we chose the case of $t=10$ to study the extrapolations taking into account uncertainties. The previous fits were performed using the central values of the $\Xi N$ lattice QCD phase shifts with $32 \leq E_{\rm{cm}}\leq 32.8$ MeV. We have also fitted to the upper bound and lower bound of the lattice QCD results of the $\Xi N$ channel\footnote{Please refer to Fig.~2(b) of Ref.~\cite{Sasaki:2018mzh}.}. The near-threshold $\Xi N$ phase shifts at $m_\pi =146$ MeV, at the physical point and the extrapolations of $i/a_{\Xi N}$ for all the three cases are shown in Fig.~\ref{PP10Er}. These results show that as the $\Xi N$ $^1S_0$ phase shifts at $m_{\pi}=146$ MeV decrease, the
slope of the trajectories with the extrapolation becomes smaller. In particular, if we use the lower bound of the $\Xi N$ lattice QCD phase shifts at $t=10$, the quasi-bound state survives in the physical region.

\subsection{Cross sections and low energy parameters}

\begin{figure}[h]
 \centering
	\includegraphics[width=0.5\textwidth,bb=0 0 286 211]{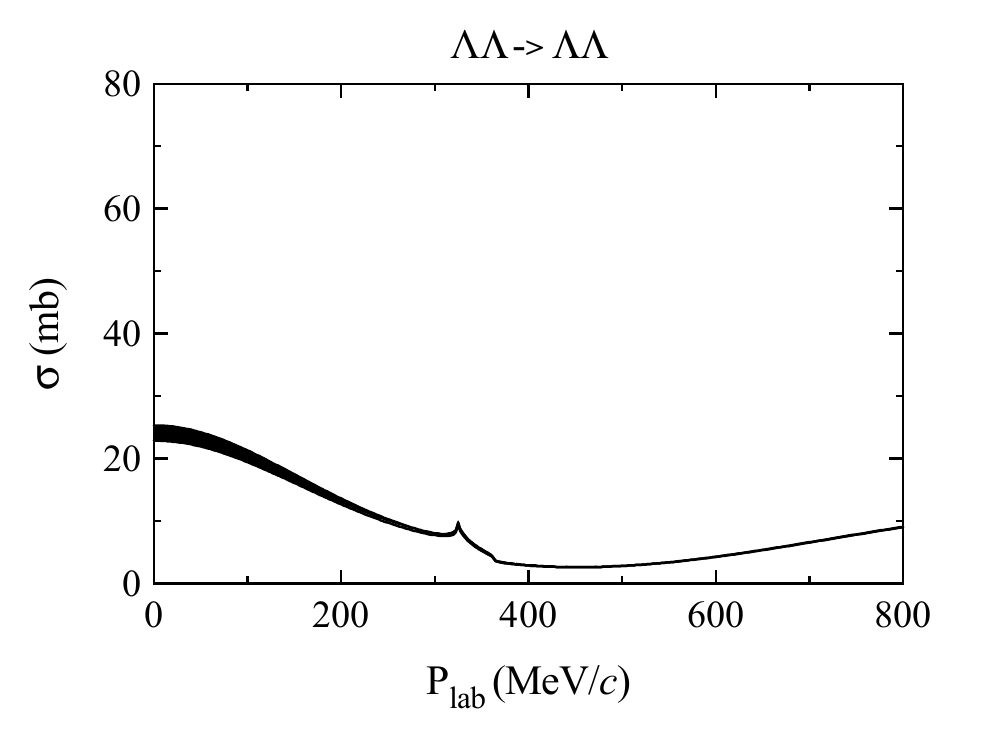}~
	\includegraphics[width=0.5\textwidth,bb=0 0 286 211]{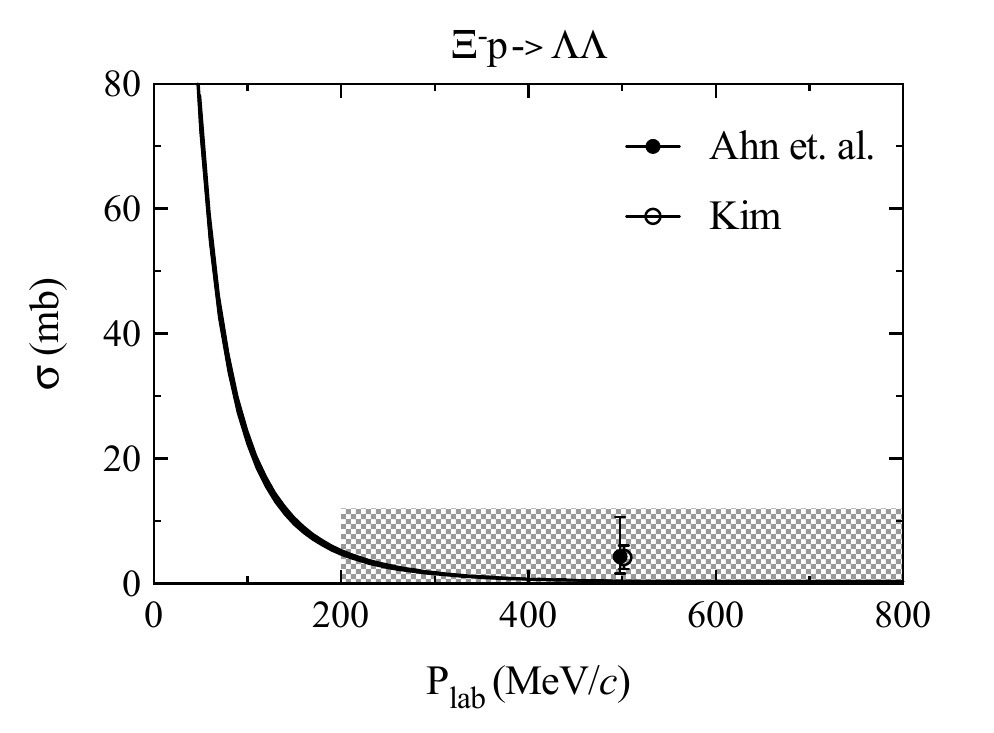} \\
	\includegraphics[width=0.5\textwidth,bb=0 0 286 211]{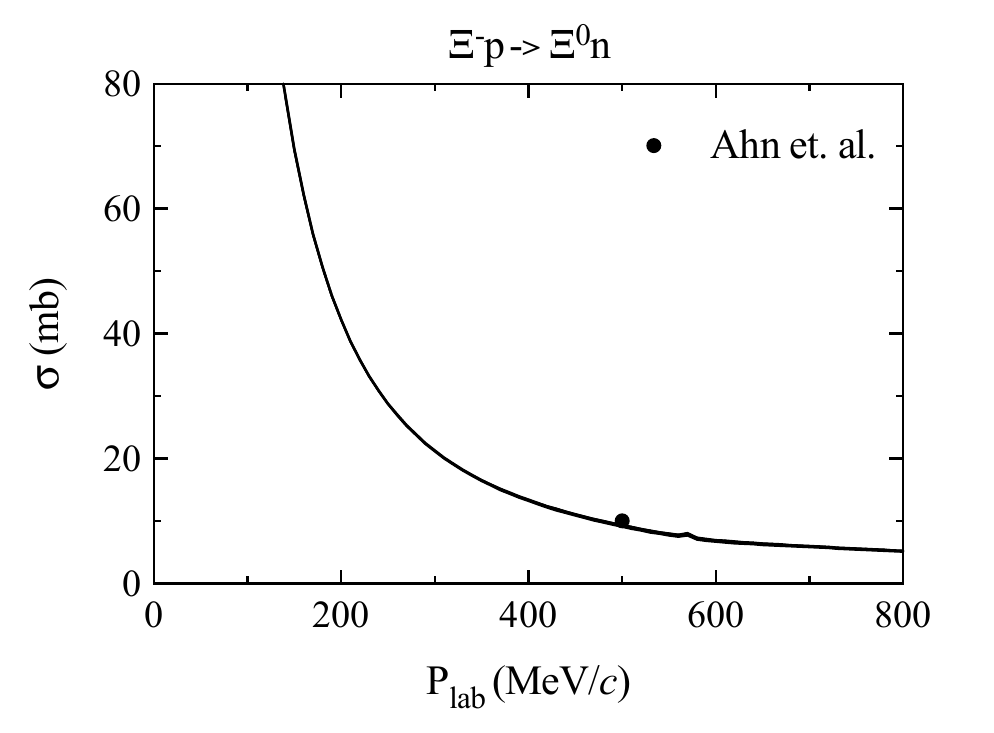}~
	\includegraphics[width=0.5\textwidth,bb=0 0 286 211]{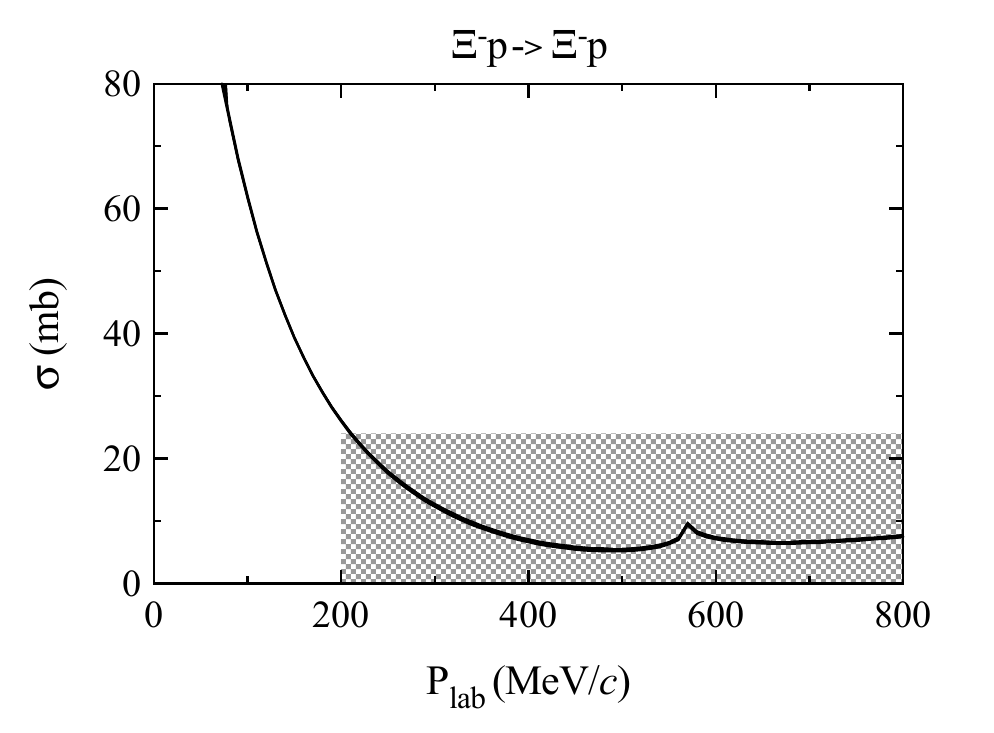}
	\caption{$\Lambda\Lambda$ and $\Xi^-p$ induced cross sections with the LECs obtained by fitting to the $t=10$ lattice QCD data. The experimental data are taken from Refs.~\cite{Ahn:2005jz,Kim:2015hyp}. The grid bands in $\Xi^-p \rightarrow \Lambda\Lambda$ and $\Xi^-p \rightarrow \Xi^- p$ reactions show the upper limits from Ref.~\cite{Ahn:2005jz}. }\label{cs}
\end{figure}

Finally we compare our results with the available experimental data. 
We note that the cross sections calculated using the relevant LECs determined with the $t=9$ and $t=10$ lattice QCD data are more consistent with their experimental counterparts than those obtained with the 
$t=11$ and $t=12$ lattice QCD data. Following the preceding paragraph, we study the case of  $t=10$ in this sector.
In Fig.~\ref{cs}, we show the $\Lambda\Lambda$ and $\Xi^-p$ induced cross sections with the statistical errors discussed previously taken into account. The cross sections are calculated with all the partial waves with total angular momentum $J\leq 2$. The experimental data are taken from Refs.~\cite{Ahn:2005jz,Kim:2015hyp}. One can see that our results are consistent with the scattering data, although the latter has a sizable uncertainty. Such a comparison shows that 
the lattice QCD data (in particular, those obtained with $t=10$), the relativistic ChEFT approach and the experimental data are in general consistent with each other. 

In Table.~\ref{allSL},  we summarize the scattering lengths and effective ranges  for various channels with the LECs determined by fitting to the  $t=10$ lattice QCD data. 
For the sake of comparison, we show as well
 the next-to-leading order (NLO) and LO~\cite{Polinder:2007mp} heavy baryon (HB) ChEFT~\cite{Haidenbauer:2015zqb} results obtained with a cutoff $\Lambda_F=600$ MeV, those of the NSC97f model~\cite{Stoks:1999bz} and the fss2 model~\cite{Fujiwara:2006yh}. Note that the Coulomb force is considered in the latter two approaches.  The results from different approaches are rather scattered. Clearly, more experimental information are needed to further constrain
the $S=-2$ baryon-baryon interactions.

It is interesting to compare our results with those of the NLO HB approach~\cite{Haidenbauer:2015zqb}. In particular, 
  the scattering lengths of the $\Sigma^+\Sigma^+$ channel are rather different but those of the $\Lambda\Lambda$ channel are quite similar, as shown in Table~\ref{allSL}. 
  We note that in Ref.~\cite{Haidenbauer:2015zqb} they have fitted to the $pp$ phase shifts and the $\Sigma^+ p$ cross sections to fix the relevant LECs in the $S$-wave contact terms 
  with SU(3) breaking effects taken into account and then made predictions for the $\Sigma^+\Sigma^+$ channel. Our study shows that the lattice QCD data seem to prefer a weaker $\Sigma^+\Sigma^+$ attraction than
  that predicted by the NLO HB approach, indicating the suppression of the attraction as one adds more strangeness into the system may be larger than
  that considered in Ref.~\cite{Haidenbauer:2015zqb} (see also the discussion in Sec.~\ref{IVA}). Note that our results for the $\Sigma^+\Sigma^+$ channel are not dependent on $t$. On the other hand, the similar results for the $\Lambda\Lambda$ channel can be easily understood.  The NLO HB approach fixed  the relevant LECs by  the empirical value of the $\Lambda\Lambda$ scattering length within the range of $-1\sim -0.5$ fm, while our fits to the lattice QCD data also yield  a $a_{1S0}^{\Lambda\Lambda}$ consistent with its empirical value (see Table \ref{PSLL3}).


\begin{table}[h]
\centering
\caption{Predicted scattering lengths $a$ and effective ranges $r$ for various channels. The results obtained from HB ChEFT at NLO~\cite{Haidenbauer:2015zqb}, LO~\cite{Polinder:2007mp} with cutoff $\Lambda_F=600$ MeV, the NSC97f model~\cite{Stoks:1999bz} and the fss2 model~\cite{Fujiwara:2006yh} are also shown for the sake of  comparison. Note that the Coulomb force is considered in the latter two approaches.
}
\label{allSL}
\begin{tabular}{L{1.5cm}L{1.5cm}L{2.5cm}L{2.5cm}L{2.5cm}L{2.5cm}L{2cm}}
\hline
\hline
 Channel &   & This work & HB NLO~\cite{Haidenbauer:2015zqb} & HB LO~\cite{Polinder:2007mp} & NSC97f~\cite{Stoks:1999bz} & fss2~\cite{Fujiwara:2006yh} \\
\hline
$\Sigma^+\Sigma^+$ & $a_{1S0}^{\Sigma^+\Sigma^+}$ & $-0.80$ & $-1.83$ & $-7.76$& $\phantom{-}6.98$ & $-9.72$ \\
                   & $r_{1S0}^{\Sigma^+\Sigma^+}$ & ~~~$13.3$ & $\phantom{-}6.05$ & $\phantom{-}2.00$ &  $\phantom{-}1.46$ & $\phantom{-}2.26$ \\
$\Xi^0p$ & $a_{1S0}^{\Xi^0p}$ & ~~~$0.45$ & $\phantom{-}0.34$ & $\phantom{-}0.19$& $\phantom{-}0.40$ & $\phantom{-}0.33$ \\
         & $r_{1S0}^{\Xi^0p}$ & $-4.55$ & $-7.07$ & $-37.7$& $-8.94$ & $-9.23$ \\
         & $a_{3S1}^{\Xi^0p}$ & $-0.09$ & $\phantom{-}0.02$ & $\phantom{-}0.00$& $-0.03$ & $-0.20$ \\
         & $r_{3S1}^{\Xi^0p}$ & ~~~$72.5$ & ~~$1797$ & $>10^4$& ~~~~$912$ & $\phantom{-}27.4$ \\
$\Lambda\Lambda$ & $a_{1S0}^{\Lambda\Lambda}$ & $-0.60$ & $-0.66$ & $-1.52$& $-0.35$ & $-0.81$ \\
                 & $r_{1S0}^{\Lambda\Lambda}$ & ~~~$3.73$ & $\phantom{-}5.05$  & $\phantom{-}0.59$&  $\phantom{-}14.7$ & $\phantom{-}3.80$ \\
$\Xi^0n$ & $a_{3S1}^{\Xi^0n}$ & $-0.14$ & $-0.26$ & $-0.25$& &  \\
         & $r_{3S1}^{\Xi^0n}$ & $-14.0$ & $\phantom{-}5.26$ & $-8.27$& & \\
\hline
\hline
\end{tabular}

\end{table}

\section{Summary and outlook}\label{V}

Recent progress in lattice QCD simulations  provides us an unprecedented opportunity  to better understand baryon-baryon interactions that
play an important role in studies of hypernuclear  and astronuclear physics. In particular, supplementary information on hyperon-nucleon(hyperon) interactions (to scarce experimental data) are
key to understanding many important issues of current interest, such as the existence of H-, $\Omega N$, and $\Omega\Omega$ dibaryons, and the internal structure of neutron stars.  Nevertheless, present lattice QCD simulations still
suffer important systematic uncertainties originated from unphysical pion masses as well as coupled channel effects. Careful studies of such effects are urgently needed to fully utilize the
state-of-art lattice QCD simulations to advance our understanding of the non-perturbative strong interaction.

In the present work, we have studied the strangeness $S=-2$ baryon-baryon interactions in relativistic chiral effective field theory at leading order. The latest lattice QCD data of the HAL QCD Collaboration were used to fix the relevant low energy constants. We obtained a good description of the lattice QCD results (with perhaps the exception of the high energy
$I=0$ $\Xi N$ $^3 S_1$ phase shifts). Extrapolations from $m_\pi =146$ MeV to the physical region were made. The behavior of the $\Xi N$ system was found to be very  sensitive to the lattice QCD data fitted. In addition,  our results can describe the available experimental data very well,
which shows the overall consistency between lattice QCD simulations, the relativistic chiral effective field theory and the experimental data.

\section{Acknowledgements}
The authors are grateful to Kenji Sasaki for providing us the latest lattice QCD results and a careful reading of this manuscript. K. W. L. and T. H. thank Wolfram Weise for useful discussions. This work is partly supported by the National Natural Science Foundation of China under Grants No.11522539, 11735003, by the fundamental Research Funds for the Central Universities,  by JSPS KAKENHI Grants No. JP16K17694 and by the Yukawa International Program for Quark-Hadron Sciences (YIPQS).
K. W. L. acknowledges financial support from the China Scholarship Council.


\begin{thebibliography}{500}

\bibitem{Ahn:1998fj} 
  J.~K.~Ahn {\it et al.} [KEK-PS E224 Collaboration],
  Phys.\ Lett.\ B {\bf 444}, 267 (1998).
  
\bibitem{Sasaki:2018mzh} 
  K.~Sasaki {\it et al.} [HAL QCD Collaboration],
  EPJ Web Conf.\  {\bf 175}, 05010 (2018).
  
\bibitem{Jaffe:1976yi} 
  R.~L.~Jaffe,
  Phys.\ Rev.\ Lett.\  {\bf 38}, 195 (1977)
  Erratum: [Phys.\ Rev.\ Lett.\  {\bf 38}, 617 (1977)].
  
\bibitem{Beane:2010hg} 
  S.~R.~Beane {\it et al.} [NPLQCD Collaboration],
  Phys.\ Rev.\ Lett.\  {\bf 106}, 162001 (2011)
  [arXiv:1012.3812 [hep-lat]].
  
\bibitem{Inoue:2010es} 
  T.~Inoue {\it et al.} [HAL QCD Collaboration],
  Phys.\ Rev.\ Lett.\  {\bf 106}, 162002 (2011)
  [arXiv:1012.5928 [hep-lat]].
  
\bibitem{Beane:2011iw} 
  S.~R.~Beane {\it et al.} [NPLQCD Collaboration],
  Phys.\ Rev.\ D {\bf 85}, 054511 (2012)
  [arXiv:1109.2889 [hep-lat]].
  
\bibitem{Francis:2018qch} 
  A.~Francis, J.~R.~Green, P.~M.~Junnarkar, C.~Miao, T.~D.~Rae and H.~Wittig,
  arXiv:1805.03966 [hep-lat].
    
\bibitem{Shanahan:2011su} 
  P.~E.~Shanahan, A.~W.~Thomas and R.~D.~Young,
  Phys.\ Rev.\ Lett.\  {\bf 107}, 092004 (2011)
  [arXiv:1106.2851 [nucl-th]].
  
\bibitem{Haidenbauer:2011ah} 
  J.~Haidenbauer and U.~G.~Meissner,
  Phys.\ Lett.\ B {\bf 706}, 100 (2011)
  [arXiv:1109.3590 [hep-ph]].
  
\bibitem{Inoue:2011ai} 
  T.~Inoue {\it et al.} [HAL QCD Collaboration],
  Nucl.\ Phys.\ A {\bf 881}, 28 (2012)
  [arXiv:1112.5926 [hep-lat]].

\bibitem{Yamaguchi:2016kxa} 
  Y.~Yamaguchi and T.~Hyodo,
  Phys.\ Rev.\ C {\bf 94}, 065207 (2016)
  [arXiv:1607.04053 [hep-ph]].

\bibitem{HALQCD:2012aa} 
  N.~Ishii {\it et al.} [HAL QCD Collaboration],
  Phys.\ Lett.\ B {\bf 712}, 437 (2012)
  [arXiv:1203.3642 [hep-lat]].
  
\bibitem{Aoki:2012tk} 
  S.~Aoki {\it et al.} [HAL QCD Collaboration],
  PTEP {\bf 2012}, 01A105 (2012)
  [arXiv:1206.5088 [hep-lat]].
  
\bibitem{Yamaguchi:2001ip} 
  M.~Yamaguchi, K.~Tominaga, T.~Ueda and Y.~Yamamoto,
  Prog.\ Theor.\ Phys.\  {\bf 105}, 627 (2001).
  
\bibitem{Friedman:2007zza} 
  E.~Friedman and A.~Gal,
  Phys.\ Rept.\  {\bf 452}, 89 (2007)
  [arXiv:0705.3965 [nucl-th]].
  
\bibitem{Hiyama:2010zz} 
  E.~Hiyama, M.~Kamimura, Y.~Yamamoto, T.~Motoba and T.~A.~Rijken,
  Prog.\ Theor.\ Phys.\ Suppl.\  {\bf 185}, 152 (2010).
  
\bibitem{Khaustov:1999bz} 
  P.~Khaustov {\it et al.} [AGS E885 Collaboration],
  Phys.\ Rev.\ C {\bf 61}, 054603 (2000)
  [nucl-ex/9912007].
  
\bibitem{Kohno:2009ny} 
  M.~Kohno and S.~Hashimoto,
  Prog.\ Theor.\ Phys.\  {\bf 123}, 157 (2010)
  [arXiv:0909.1385 [nucl-th]].
  
\bibitem{Krishichayan:2010zza} 
  Krishichayan, X.~Chen, Y.-W.~Lui, Y.~Tokimoto, J.~Button and D.~H.~Youngblood,
  Phys.\ Rev.\ C {\bf 81}, 014603 (2010).
  
\bibitem{Nakazawa:2015joa} 
  K.~Nakazawa {\it et al.},
  PTEP {\bf 2015}, 033D02 (2015).
  
\bibitem{Garcilazo:2015noa} 
  H.~Garcilazo and A.~Valcarce,
  Phys.\ Rev.\ C {\bf 92}, 014004 (2015)
  [arXiv:1507.03733 [hep-ph]].
  
\bibitem{Demorest:2010bx} 
  P.~Demorest, T.~Pennucci, S.~Ransom, M.~Roberts and J.~Hessels,
  Nature {\bf 467}, 1081 (2010)
  [arXiv:1010.5788 [astro-ph.HE]].
  
\bibitem{Antoniadis:2013pzd} 
  J.~Antoniadis {\it et al.},
  Science {\bf 340}, 6131 (2013)
  [arXiv:1304.6875 [astro-ph.HE]].
  
\bibitem{Lonardoni:2013gta} 
  D.~Lonardoni, F.~Pederiva and S.~Gandolfi,
  Phys.\ Rev.\ C {\bf 89}, 014314 (2014)
  [arXiv:1312.3844 [nucl-th]].
  
\bibitem{Maslov:2015msa} 
  K.~A.~Maslov, E.~E.~Kolomeitsev and D.~N.~Voskresensky,
  Phys.\ Lett.\ B {\bf 748}, 369 (2015)
  [arXiv:1504.02915 [astro-ph.HE]].
  
\bibitem{Ren:2016jna} 
  X.~L.~Ren, K.~W.~Li, L.~S.~Geng, B.~W.~Long, P.~Ring and J.~Meng,
  Chin.\ Phys.\ C {\bf 42}, 014103 (2018)
  [arXiv:1611.08475 [nucl-th]].
  
\bibitem{Ren:2017yvw} 
  X.~L.~Ren, K.~W.~Li, L.~S.~Geng and J.~Meng,
  arXiv:1712.10083 [nucl-th].
  
\bibitem{Li:2016paq} 
  K.~W.~Li, X.~L.~Ren, L.~S.~Geng and B.~Long,
  Phys.\ Rev.\ D {\bf 94}, 014029 (2016)
  [arXiv:1603.07802 [hep-ph]].
  
\bibitem{Li:2016mln} 
  K.~W.~Li, X.~L.~Ren, L.~S.~Geng and B.~W.~Long,
  Chin.\ Phys.\ C {\bf 42}, 014105 (2018)
  [arXiv:1612.08482 [nucl-th]].
  
\bibitem{Li:2017zwn} 
  K.~W.~Li, X.~L.~Ren, L.~S.~Geng and B.~Long,
  PoS INPC {\bf 2016}, 276 (2017)
  [arXiv:1701.06272 [nucl-th]].
  
\bibitem{Song:2018qqm} 
  J.~Song, K.~W.~Li and L.~S.~Geng,
  Phys.\ Rev.\ C {\bf 97}, 065201 (2018)
  [arXiv:1802.04433 [nucl-th]].
  
\bibitem{Ren:2018xxd} 
  X.~L.~Ren, K.~W.~Li and L.~S.~Geng,
  Nucl.\ Phys.\ Rev.\  {\bf 34}, 392 (2017)
  [arXiv:1801.00844 [nucl-th]].

\bibitem{Ren:2017thl} 
  X.~L.~Ren, K.~W.~Li and L.~S.~Geng,
  arXiv:1709.10266 [nucl-th].
  
  
\bibitem{Epelbaum:2008ga} 
  E.~Epelbaum, H.~W.~Hammer and U.~G.~Meissner,
  Rev.\ Mod.\ Phys.\  {\bf 81}, 1773 (2009)
  [arXiv:0811.1338 [nucl-th]].
  
\bibitem{Machleidt:2011zz} 
  R.~Machleidt and D.~R.~Entem,
  Phys.\ Rept.\  {\bf 503}, 1 (2011)
  [arXiv:1105.2919 [nucl-th]].
  
\bibitem{Haidenbauer:2013oca} 
  J.~Haidenbauer, S.~Petschauer, N.~Kaiser, U.-G.~Meissner, A.~Nogga and W.~Weise,
  Nucl.\ Phys.\ A {\bf 915}, 24 (2013)
  [arXiv:1304.5339 [nucl-th]].

\bibitem{Polinder:2007mp} 
  H.~Polinder, J.~Haidenbauer and U.-G.~Meissner,
  Phys.\ Lett.\ B {\bf 653}, 29 (2007)
  [arXiv:0705.3753 [nucl-th]].

\bibitem{Haidenbauer:2015zqb} 
  J.~Haidenbauer, U.~G.~Meissner and S.~Petschauer,
  Nucl.\ Phys.\ A {\bf 954}, 273 (2016)
  [arXiv:1511.05859 [nucl-th]].
  
\bibitem{Haidenbauer:2009qn} 
  J.~Haidenbauer and U.-G.~Meissner,
  Phys.\ Lett.\ B {\bf 684}, 275 (2010)
  [arXiv:0907.1395 [nucl-th]].
  
\bibitem{Weinberg:1990rz} 
  S.~Weinberg,
  Phys.\ Lett.\ B {\bf 251}, 288 (1990).
  
\bibitem{Weinberg:1991um} 
  S.~Weinberg,
  Nucl.\ Phys.\ B {\bf 363}, 3 (1991).
  
\bibitem{private}
Private discussion with Kenji Sasaki.



  
\bibitem{Rijken:2006ep} 
  T.~A.~Rijken and Y.~Yamamoto,
  Phys.\ Rev.\ C {\bf 73}, 044008 (2006)
  [nucl-th/0603042].
  
\bibitem{Fujiwara:2006yh} 
  Y.~Fujiwara, Y.~Suzuki and C.~Nakamoto,
  Prog.\ Part.\ Nucl.\ Phys.\  {\bf 58}, 439 (2007)
  [nucl-th/0607013].
  
\bibitem{Filikhin:2002wm} 
  I.~N.~Filikhin and A.~Gal,
  Nucl.\ Phys.\ A {\bf 707}, 491 (2002)
  [nucl-th/0203036].
  
\bibitem{Afnan:2003ty} 
  I.~R.~Afnan and B.~F.~Gibson,
  Phys.\ Rev.\ C {\bf 67}, 017001 (2003).
  
\bibitem{Filikhin:2004sn} 
  I.~Filikhin, A.~Gal and V.~M.~Suslov,
  Nucl.\ Phys.\ A {\bf 743}, 194 (2004)
  [nucl-th/0406049].
  
\bibitem{Yamada:2004ks} 
  T.~Yamada,
  Phys.\ Rev.\ C {\bf 69}, 044301 (2004)
  [nucl-th/0403016].
  
\bibitem{Vidana:2003ic} 
  I.~Vidana, A.~Ramos and A.~Polls,
  Phys.\ Rev.\ C {\bf 70}, 024306 (2004)
  [nucl-th/0307096].
  
\bibitem{Usmani:2004vs} 
  Q.~N.~Usmani, A.~R.~Bodmer and B.~Sharma,
  Phys.\ Rev.\ C {\bf 70}, 061001 (2004).
  
\bibitem{Nemura:2004xb} 
  H.~Nemura, S.~Shinmura, Y.~Akaishi and K.~S.~Myint,
  Phys.\ Rev.\ Lett.\  {\bf 94}, 202502 (2005)
  [nucl-th/0407033].
  
\bibitem{Gasparyan:2011kg} 
  A.~M.~Gasparyan, J.~Haidenbauer and C.~Hanhart,
  Phys.\ Rev.\ C {\bf 85}, 015204 (2012)
  [arXiv:1111.0513 [nucl-th]].
  
\bibitem{Morita:2014kza} 
  K.~Morita, T.~Furumoto and A.~Ohnishi,
  Phys.\ Rev.\ C {\bf 91}, no. 2, 024916 (2015)
  [arXiv:1408.6682 [nucl-th]].
  
\bibitem{Adamczyk:2014vca} 
  L.~Adamczyk {\it et al.} [STAR Collaboration],
  Phys.\ Rev.\ Lett.\  {\bf 114}, no. 2, 022301 (2015)
  [arXiv:1408.4360 [nucl-ex]].
  
  
  
  
  
  
\bibitem{Ahn:2005jz} 
  J.~K.~Ahn {\it et al.},
  Phys.\ Lett.\ B {\bf 633}, 214 (2006)
  [nucl-ex/0502010].

\bibitem{Kim:2015hyp} 
  S.J. Kim, 
  presentation at the 12th International Conference on Hypernuclear and Strange Particle Physics, Sendai, Japan, see http://lambda.phys.tohoku.ac.jp/hyp2015/, 2015.    


\bibitem{Stoks:1999bz} 
  V.~G.~J.~Stoks and T.~A.~Rijken,
  Phys.\ Rev.\ C {\bf 59}, 3009 (1999)
  [nucl-th/9901028].
  
  

\end{thebibliography}
\end{document}